\documentclass{emulateapj}

\usepackage{amssymb}
\usepackage{amsmath}
\usepackage{natbib}
\usepackage{txfonts}
\usepackage[colorlinks=true,citecolor=blue,linkcolor=blue]{hyperref}
\usepackage{microtype}
\usepackage{color}
\usepackage{graphicx}
\usepackage{epstopdf}

\def\apjl{ApJL}
\def\apjs{ApJS}

\def\aap{A\&A}

\def\ufabc{1}
\def\soton{2}
\def\ufma{3}

\shorttitle{Phase transition effects on the dynamical stability of hybrid neutron stars}
\shortauthors{Pereira, Flores and Lugones}

\begin{document}

\title{Phase transition effects on the dynamical stability of hybrid neutron stars}

\author{Jonas P.~Pereira\altaffilmark{\ufabc,\soton}, C\'esar V. Flores\altaffilmark{\ufabc,\ufma} and Germ\'an Lugones\altaffilmark{\ufabc}}

\altaffiltext{\ufabc}{Universidade Federal do ABC, Centro de Ci\^encias Naturais e Humanas, Avenida dos Estados, 5001- Bang\'u, CEP 09210-170, Santo Andr\'e, SP, Brazil}

\altaffiltext{\soton}{Mathematical Sciences and STAG Research Centre,
University of Southampton, Southampton, SO17 1BJ, United Kingdom}

\altaffiltext{\ufma}{Universidade Federal do  Maranh\~ao, Departamento de F\'isica, Campus Universit\'ario do Bacanga, CEP 65080-805, S\~ao Lu\'is, Maranh\~ao, Brazil.}

\begin{abstract}

We study radial oscillations of hybrid non-rotating neutron stars composed by a quark matter core and hadronic external layers. At first, we physically deduce the junction conditions that should be imposed between two any phases in these systems when perturbations take place. Then we compute the oscillation spectrum focusing on the effects of slow and rapid phase transitions at the quark-hadron interface.  We use a generic MIT bag model for quark matter and  a relativistic mean field theory for hadronic matter.  In the case of rapid transitions at the interface we find a general relativistic version of the reaction mode which has similar properties as its classical counterpart. We also show that the usual static stability condition $\partial M/\partial \rho_c\geq 0$, where $\rho_c$ is the central density of a star whose total mass is $M$, remains always true for rapid transitions but breaks down in general for slow transitions. In fact, for slow transitions we find that the frequency of the fundamental mode can be a real number (indicating stability) even for some branches of stellar models that verify $\partial M/\partial \rho_c \leq 0$.  Thus, when  secular instabilities are suppressed, as expected below some critical stellar rotation rate, it would be possible the existence of twin or even triplet stars with the same gravitational mass but different radii, with one of the counterparts having $\partial M/\partial \rho_c \leq 0$. We explore some astrophysical consequences of these results. 

\end{abstract}


\altaffiltext{}{jonas.pereira@ufabc.edu.br; cesarovfsky@gmail.com; german.lugones@ufabc.edu.br}

\maketitle

\section{Introduction}

It is known since long ago that the stability of one-phase stars  can be assessed by means of families of static solutions with different central densities $\rho_c$, such that unstable ones have $\partial M/\partial \rho_c<0$ \citep{Wheeler1965,1986bhwd.book.....S}. However, it has been shown in several different scenarios (charged strange quark stars \citep{2015PhRvD..92h4009A}, hybrid stars with a quark and hadronic phases \citep{2012IJMPS..18..105V}, color superconducting quark stars \citep{2010PhRvD..82f3006V}, etc.) that this simple condition may not hold in general. 
Studies on multi-phase neutron stars are pertinent because they could unveil characteristics uniquely pertaining to these stars and thus lead to potential observables for their probe. Of special interest, due to stability assessments and motivated by the advent of gravitational wave astronomy \citep{2016PhRvL.116f1102A,2016PhRvL.116x1103A,PhysRevLett.118.221101,2017PhRvL.119p1101A}, is the eigenfrequency spectrum of these systems. This is very important because it carries information  of the stars' structures, even their hybrid nature \citep{2014CQGra..31o5002V}. The most natural approach thereof is by means of radial perturbations, the route we take in this work. 

As regards hybrid stars' perturbations, in general one cannot integrate the system of equations as is done in one-phase stars, but must instead take due care of additional boundary conditions at the phase-splitting interfaces,     which in essence capture the main physics taking place in their vicinities. There are basically two kinds of physical behavior around such surfaces in the presence of perturbations: either volume elements are converted from one phase to the other or they keep their nature and are only stretched or compressed. The first kind of conversion is related to rapid phase transitions, while the other regards slow phase transitions \citep{1989A&A...217..137H}. 
Suggestions of possible physics behind them include quantum and thermal nucleation processes  of one phase into the other that result in characteristic conversion times that are very sensitive to several microphysical details of both phases (see e.g. \cite{Lugones:2011xv,Bombaci:2016xuj,Lugones:2015bya,Kroff:2014qxa} and references therein). As a consequence, the transition might be either rapid or slow, which motivates the investigations performed in this paper. More details about the aforesaid phase conversions will be given in the subsequent sections.

There are many studies related to radial oscillations of hybrid stars, focusing for instance on the influence of the mixed phase  \citep{2002ApJ...579..374G,2002ApJ...566L..89S}, electric charge \citep{2014EL....10539001B}, the observational consequences of a possible quark core formation (mixed phase is taken into account) \citep{2003PhLB..552....1M}, as well as pion condensation \citep{1979PhLB...83..158M,1982ApJ...258..306H}. Puzzling enough, it seems that rapid phase transitions and its consequences for hybrid stars have not been investigated much in the literature.  To the best of our knowledge it has been pioneered in the context of Newtonian stars by Haensel and collaborators \citep{1989A&A...217..137H}, who found the so called reaction mode, but general relativistic analyses seem yet pending. 
Additionally, a systematic analysis of stellar stability in the case of slow phase transitions has not yet been performed in the literature.
We try to partially fill this gap here.  

The strategy of this work is as follows. In the next three sections we summarize the equations of state and the relevant equations concerning radial stability analyses. In Sec. \ref{junction} we deduce the relevant boundary conditions for matching the phases of hybrid stars when perturbations take place. Section \ref{numerical} is devoted to a brief explanation of the numerical approaches we have made use of in order to obtain our results, present in Sec. \ref{results}.  Finally, in Sec. \ref{discussion} we discuss the main issues raised in our analyses. We work with geometric units unless otherwise stated.

\section{Equations of state}
\label{section_eos}

In this work we investigate the imprint of different phase transitions into the stability of hybrid stars having quark and hadronic phases.
Given the unknown constitution of matter at very high densities, the equation of state (EOS) is usually derived from phenomenological descriptions trying to take into account the expected properties of matter at each density regime.

\subsection{Hadronic Matter}

For the hadronic phase we use a non-linear Walecka model \citep{Walecka:1974qa,glendenning1991}
including the whole baryon octet, electrons and the corresponding antiparticles. The Lagrangian is given by
\begin{equation}
{\cal L}={\cal L}_{B}+{\cal L}_{M}+{\cal L}_{L}, \label{octetlag}
\end{equation}
where the indices $B$, $M$ and $L$ refer to baryons, mesons and leptons respectively. 
The only baryons considered here are nucleons (neutrons and protons), thus ${\cal L}_B$ reads:
\begin{eqnarray}
{\cal L}_{B=n, p} &=&  \sum_B \bar \psi_B [\gamma^\mu\left
(i\partial_\mu - g_{\omega} \omega_\mu- g_{\rho} \ \vec \tau
\cdot \vec \rho_\mu \right)\nonumber \\  && - (m-g_{\sigma} \sigma)]\psi_B.
\end{eqnarray}
The contribution of the mesons $\sigma$, $\omega$ and $\rho$ is given by
\begin{eqnarray}
{\cal L}_{M} &=& \tfrac{1}{2} (\partial_{\mu} \sigma \ \!
\partial^{\mu}\sigma -m_\sigma^2 \ \! \sigma^2) - \tfrac{b}{3} \ \!
m_N\ \! (g_\sigma\sigma)^3 -\tfrac{c}{4} \ (g_\sigma \sigma)^4
\nonumber\\
& & -\tfrac{1}{4}\ \omega_{\mu\nu}\ \omega^{\mu\nu} +\tfrac{1}{2}\
m_\omega^2 \ \omega_{\mu}\ \omega^{\mu} \nonumber \\     && -\tfrac{1}{4}\ \vec \rho_{\mu\nu} \cdot \vec \rho\ \! ^{\mu\nu}+
\tfrac{1}{2}\ m_\rho^2\  \vec \rho_\mu \cdot \vec \rho\ \! ^\mu.
\end{eqnarray}
Electrons are included as a free Fermi gas, ${\cal L}_{L}=\sum_l \bar \psi_l \left(i \rlap/\partial -
m_l\right)\psi_l$, in chemical equilibrium under weak interactions with all other particles.

The constants in the model are determined by the properties of nuclear matter and hyperon potential depths known from hypernuclear experiments. In the present work we use the NL3 parametrization for which we have $(g_{\sigma}/m_{\sigma})^{2} = 15.8$ fm$^{-2}$, $(g_{\omega}/m_{\omega})^{2} = 10.51$ fm$^{-2}$, $(g_{\rho}/m_{\rho})^{2} = 5.35$ fm$^{-2}$, $b = 0.002055$ and $c = -0.002651$  \citep{1997PhRvC..55..540L,glendenning1991}.  At low densities we use the Baym, Pethick and Sutherland (BPS) model \citep{bps}.
For details on the explicit form of the equation of state derived from this Lagrangian the reader is referred to \cite{lugones_eos1,lugones_eos2} and references therein.


\subsection{Quark Matter}

For quark matter, bag-like models are appealing due to their simplicity and ability to effectively capture some key aspects of QCD. Here, we use the bag-like model presented in \cite{2005ApJ...629..969A}, with free parameters $B$, $a_4$ and $a_2$, which could encompass several physical effects, as explained below. The model is defined by the following grand thermodynamic potential 
\begin{equation}
\Omega =  -\frac{3}{4\pi^2} a_4 \mu^4 + \frac{3}{4\pi^2}a_2\mu^2 + B + \Omega_e
\label{grand_potential},
\end{equation}
where $\mu \equiv (\mu_u+\mu_s+\mu_d)/3$ is the quark chemical potential, and $\Omega_e$ is the grand thermodynamic potential for electrons $e$. 

For hybrid systems, though, it happens that the contribution to the thermodynamic quantities coming from electrons is negligible due to the following reason. Local charge neutrality implies that $\tfrac{2}{3} n_u - \tfrac{1}{3} n_d- \tfrac{1}{3} n_s = n_e$, where $n_i$ refers to the number density of the \textit{i}th particles involved. From the chemical equilibrium relations $\mu_d=\mu_u+\mu_e$ and $\mu_s=\mu_d$,  it follows that at zero temperature $\mu_e=m_s^2/(4\mu)-m_s^4/(48\mu^3)+O(m_s^6/\mu^5)$ \citep{2005ApJ...629..969A}, and therefore, $\mu_e/\mu \sim O(m_s^2/\mu^2)$. 
The effect of electrons in the quark matter EOS can be estimated through the ratio $\Omega_e / \Omega \sim O(\mu_e^4 /\mu^4) \sim O(m_s^8/\mu^8)$. Since we are interested in hybrid systems, the quark matter EOS is used essentially in the high density regime where $\mu$ is significantly larger than $m_s$ (typically, $\mu \sim 300$ MeV and $m_s \sim 100$ MeV),  which means a very small contribution of the electrons to the quark EOS and hence to the star's macroscopic parameters.  This is in agreement with  other estimates in the literature (see e.g. page 368 of \cite{haensel2007neutron} and references therein).  Therefore, we neglect the contribution of electrons in Eq. (\ref{grand_potential}), which has the clear advantage of leading to analytic thermodynamic expressions. 
 
As explained in \cite{2005ApJ...629..969A}, the above phenomenological model is convenient because, besides the standard MIT-bag model ($a_4=1, a_2=m_s^2$), it also allows exploring the effect of strong interactions (by modifying $a_4$), as well as the effects of color superconductivity ($a_2=m_s^2-4\Delta^2$, being $\Delta$ the energy gap associated with the quark pairing). 

From Eq. (\ref{grand_potential}), one can obtain the pressure $p = - \Omega$, the baryon number density 
\begin{equation}
n_b = -\frac{1}{3}\frac{\partial\Omega}{\partial \mu}=\frac{1}{2\pi^2} (2a_4 \mu^3 - a_2\mu )
\label{nbquarks},
\end{equation}
and the energy density
\begin{equation}
\epsilon=3\mu n_b-p= \frac{9}{4\pi^2}a_4\mu^4 - \frac{3}{4\pi^2} a_2\mu^2 + B
\label{energy_dens}.
\end{equation}
An analytic expression of the form $p = p(\epsilon)$  for this phenomenological model can be simply obtained if one solves Eq. (\ref{energy_dens}) for $\mu$ and replaces it in Eq. (\ref{grand_potential}). The final result is
\begin{equation}
p(\epsilon)= \tfrac{1}{3}(\epsilon - 4B)-\frac{a_2^2}{12\pi^2a_4}\left[1+\sqrt{1+\frac{16\pi^2a_4}{a_2^2}(\epsilon -B)}\label{eos_qm} \right].
\end{equation}

\begin{figure}[tb]
    \centering
    \includegraphics[width=\columnwidth]{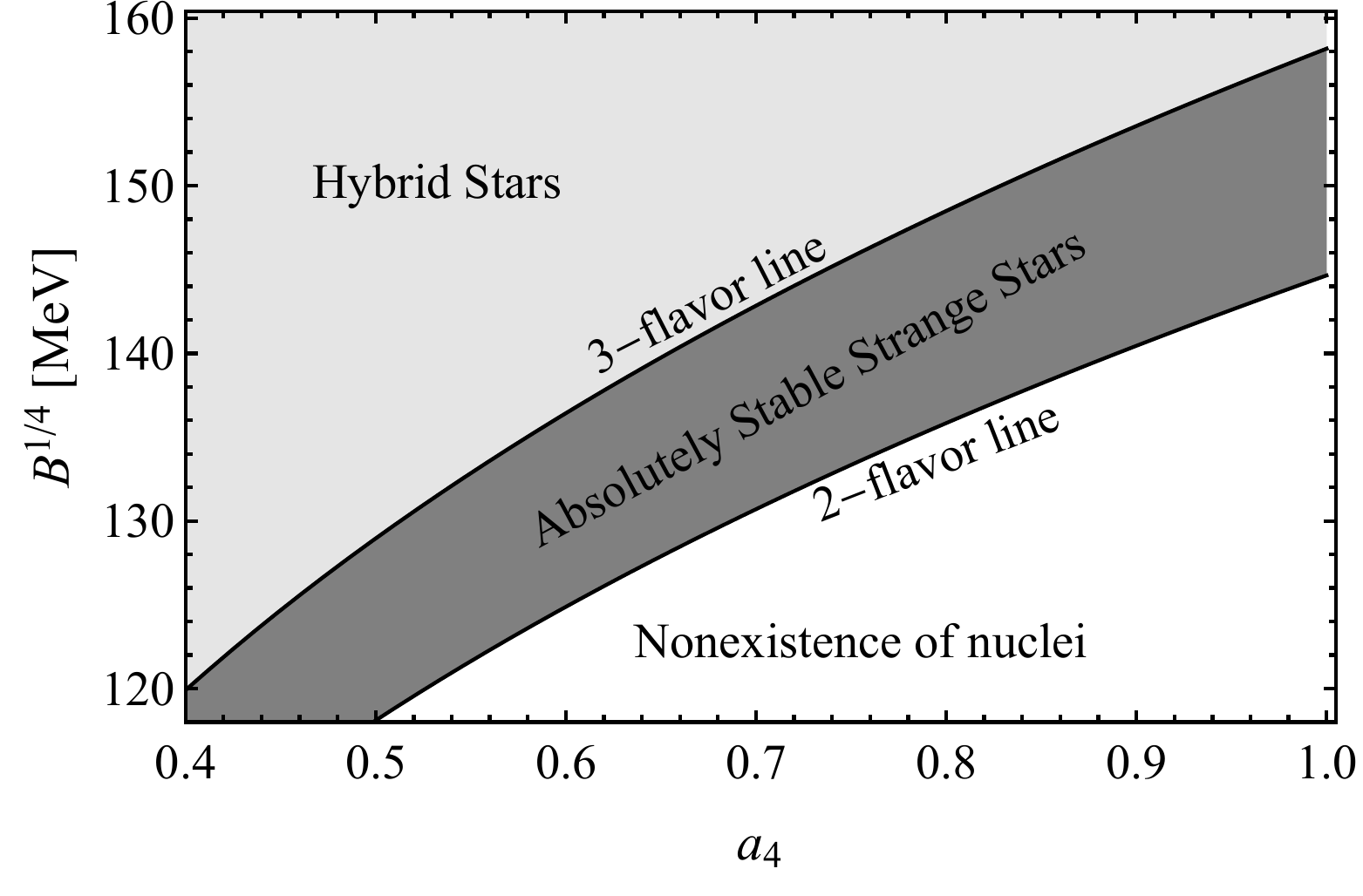}
    \caption{Regions in the $(B^{1/4},a_4)$ plane (for $a_2^{1/2}=100$ MeV) associated with hybrid neutron stars and absolutely stable strange quark stars. Parameters in the region below the 2-flavor line are unphysical due to the known existence of nuclei. } 
\label{hyb_star_stable_QS}
\end{figure}

Nevertheless, it is not for all values of $a_4$ and the effective bag constant $B$ that hybrid stars exist. 
They only do if the phase transition pressure is larger than zero. In this work we assume that the interface between hadrons and quarks is a sharp discontinuity (justifications for it are given in Sec. \ref{discussion}) at which the Gibbs conditions $p_{quarks} = p_{hadrons}$ and  $g_{quarks} = g_{hadrons}$ are satisfied, being $g=(p+\epsilon)/n_b=3\mu$ the Gibbs free energy per baryon \citep{1986bhwd.book.....S}. 
Thus, the quark-hadron interface will be located at $p > 0$ only if  $B>B_{min}$, with
\begin{equation}
B_{min}=\frac{g^2(0)}{108 \pi^2}[g^2(0)a_4-9a_2]\label{B_min_hyb_stars},
\end{equation}
where $g(0) = \epsilon/n_b$ is the Gibbs free energy per baryon of quark matter evaluated at null pressure. By the Gibbs condition, $g(0)$ turns out to be exactly the energy per baryon of pressureless hadronic matter, taken here to be that of the iron, of approximately $930$ MeV. The above condition  defines the 3-flavor line in Fig. \ref{hyb_star_stable_QS}. 

One should also guarantee that the hadronic part of a hybrid star is not in metastable equilibrium. This is done by imposing that the energy per baryon of $ud$ quark matter is larger than the iron binding energy. For two-flavor quark matter in the absence of electrons and in the massless limit one has that
\begin{equation}
\Omega_{2 \mathrm{f}}=-\tilde{p}=-\frac{24a_4}{4\pi^2(1+2^{1/3})^3}\tilde{\mu}^4+\frac{2a_2}{4\pi^2}\tilde{\mu}^2+B\label{2-flavor-omega},
\end{equation}
where $\tilde{\mu}\equiv (\mu_u+\mu_d)/2=(1+2^{1/3})\mu_u$ (due to local charge neutrality). From $\tilde{n}_b= -\frac{1}{3} (\partial \Omega_{2\mathrm{f}}/\partial \tilde{\mu})$ and $\tilde{\epsilon}=-\tilde{p}+n_u\mu_u+n_d\mu_d=-\tilde{p}+3\tilde{n}_b\tilde{\mu}$, one has that the aforesaid 2-flavor stability condition leads to $B>\tilde{B}_{min}$, where
\begin{equation}
\tilde{B}_{min}= \frac{g^2(0)}{54\pi^2}\left[\frac{4g^2(0)a_4}{(1+2^{1/3})^3}-3a_2\right]\label{B_min_2-f},
\end{equation}
which defines the 2-flavor line in Fig. \ref{hyb_star_stable_QS}. 

Fig. \ref{hyb_star_stable_QS} depicts the regions in the $(B^{1/4},a_4)$ plane (for a specific value of $a_2$) associated with hybrid neutron stars and absolutely stable strange quark stars. 
For model parameters above the 3-flavor line, stars containing quark matter must be hybrid.  
The region between the 3-flavor and 2-flavor lines defines the range of parameters related to absolutely stable strange quark stars, made up of quark matter from the center all the way up to the surface. The unshaded region of parameters (below the 2-flavor line) is excluded due to the known existence and stability of nuclei.

\subsection{Adiabatic index}

In order to solve the oscillation equations the adiabatic index
\begin{equation}
\Gamma \doteq \frac{n_b}{p} \frac{dp}{dn_b}
\end{equation}
must be obtained. 
While unperturbed stellar matter is usually assumed  to be in full thermodynamic equilibrium,  the pulsating matter may be out of equilibrium (when it is compressed and rarefied) if the relevant relaxation channels are not  quick enough with respect to the oscillation timescale. 
Therefore, a proper calculation of the adiabatic index for perturbations should take into account the (in principle) complex thermodynamic state of oscillating matter.

Note that the quark matter model presented in the previous subsection leads to an effective one-parameter equation of state, Eq. (\ref{eos_qm}), which is independent of the matter composition. Naturally, this is a consequence of having assumed the grand thermodynamic potential as given by Eq.~(\ref{grand_potential}). Since several models are included in it at once, it is ambiguous to assign the time it takes for quark matter with given $B, a_4$ and $a_2$ parameters to attain thermodynamic equilibrium upon perturbations.
Strong and Coulomb interactions have extremely short timescales and guarantee thermal equilibrium. However, chemical equilibrium depends on weak reaction timescales $\tau_{\mathrm{weak}}$ which for quark matter are known to be dependent upon many aspects of the system \citep{1993PhRvD..47..325M,1987A&A...172...95Z}.  
Depending on the temperature and the chemical potentials of the quark constituents, timescales of the order of $10^{-9}-10^ {-6}$ s might arise \citep{1993PhRvD..47..325M,Anand:1997vk}. In spite of the wide range of possible values, $\tau_{\mathrm{weak}}$ is in general much smaller than the typical timescales of perturbations in neutron stars, of the order of $10^{-4}-10^{-3}$ s.
In view of this, it seems reasonable to assume that upon perturbations, weak interactions are able to adjust particle concentrations almost instantly in such a way that the quark matter composition is always the one corresponding to chemical equilibrium. Thus, we shall consider that the adiabatic index for perturbations in quark matter is the equilibrium one ($\Gamma_{EQ}$), i.e. the same adiabatic index that determines the stiffness of the EOS in the non-pulsating configuration.

In the case of hadronic matter, there are many relaxation channels associated with strong and Coulombic elastic collisions, strong-interaction processes conserving strangeness but changing baryon fractions, modified and direct Urca processes, and nonleptonic processes \citep{2002A&A...394..213H}.  In general, the relaxation time in hadronic matter is significantly larger that the dynamical timescale of the perturbation \citep{2002A&A...394..213H}, and the value of the adiabatic index relevant for oscillations is that with fixed (frozen) composition  ($\Gamma_{FR}$) and not that calculated from the EOS of catalyzed (i.e. in full thermodynamic equilibrium) matter \citep{1966ApJ...145..514M,1977ApJ...217..799C,1995A&A...294..747G,2002A&A...394..213H}. Nonetheless, in the present work we focus on hybrid stars where hadronic matter is present mainly in a not-too-high density regime.  Model calculations \citep{2002A&A...394..213H} show that for densities below two times the nuclear saturation density ($0.32 \mathrm{fm}^{-3}$)  $\Gamma_{FR}$ and $\Gamma_{EQ}$ differ by a few percent (typically less than a 15\%).  Moreover, the real value  of the adiabatic index lies evidently between $\Gamma_{FR}$ and $\Gamma_{EQ}$.  Therefore, in view of the complications involved in an exact calculation of the adiabatic index, it is a reasonable approximation to adopt the equilibrium value $\Gamma_{EQ}$ all along the hadronic phase of the hybrid star.

\subsection{Phase conversion at the quark-hadron interface}

When a hybrid star oscillates, fluid elements around the quark-hadron interface can be compressed and rarefied upon perturbations in such a way that their pressure becomes alternatively larger and smaller than the phase transition pressure $p_t$. Since the confinement/deconfinement transition is a complex nucleation process involving not only strong interactions but also surface and curvature effects, Coulomb screening, etc. (see \cite{Lugones:2016ytl} and references therein), a fluid element that goes from a pressure below $p_t$ to a pressure above it (or vice-versa) will not necessarily undergo a phase transition. In fact, the probability of such phase conversion is associated with a nucleation timescale which at present is a model dependent quantity with an uncertain value (see \cite{Bombaci:2016xuj,Lugones:2015bya} and references therein).

If the nucleation timescale is much larger than the timescale of perturbations, we are in the case of \textit{slow} phase transitions, i.e. fluid elements keep their nature when stretched or compressed around the transition pressure. On the contrary, if the nucleation timescale is much smaller than the timescale of perturbations, we are in the case of \textit{rapid} phase transitions, i.e. fluid elements are easily converted from one phase to the other when they oscillate around the quark-hadron interface. Since the nucleation timescale is still unknown, we shall focus here on both rapid and slow phase transitions.

In a real situation, a  further comparison between $\tau_{weak}$ and the nucleation timescale could be meaningful. However, within the present analysis the properties of slow/rapid phase transitions and frozen/equilibrium oscillations depend separately on the relative value of the nucleation time with respect to the oscillation timescale, and of $\tau_{weak}$ with respect to the oscillation timescale. Thus, a similar analysis as the one presented in the previous subsection can be used to conclude that quark matter at the densest side of the interface should be in equilibrium even upon perturbations, and that hadronic matter on the other side should be treated as frozen. But, as we argued before, it is a reasonable approximation to use the equilibrium approximation everywhere in a hybrid star.

\section{Equilibrium Equations}
\label{sec3}

Before the study of radial oscillations, the compact star's equilibrium structure has to be taken into account. We consider the unperturbed (hybrid) star to be composed of layers of effective perfect fluids, whose stress-energy tensors read as
\begin{equation}
\label{tensor}
{T}_{\mu\nu} =(\epsilon + p) u_{\mu}u_{\nu} + p{g}_{\mu\nu}.
\end{equation}
At the same time, we set a static spherically symmetric background spacetime for each phase given by the following line element
\begin{equation}
\label{dsz_tov}
ds^{2}=-e^{ \nu(r)} dt^{2} + e^{ \lambda(r)} dr^{2} + r^{2}(d\theta^{2}+\sin^{2}{\theta}d\phi^{2}).
\end{equation}
After solving Einstein's equations with the definition $e^{-\lambda (r)}\doteq [1-2m(r)/r]$ we come up with the following set of stellar structure
equations (Tolman-Oppenheimer-Volkoff equations~\citep{1939PhRv...55..374O})
\begin{eqnarray}
\label{tov1}
&&\frac{dp}{dr} = - \frac{\epsilon m}{r^2}\bigg(1 + \frac{p}{\epsilon}\bigg)
	\bigg(1 + \frac{4\pi p r^3}{m}\bigg)\bigg(1 -
	\frac{2m}{r}\bigg)^{-1},
\\ \nonumber \\
\label{tov2}
&&\frac{d\nu}{dr} = - \frac{2}{\epsilon + p} \frac{dp}{dr},
\\ \nonumber \\
\label{tov3}
&&\frac{dm}{dr} = 4 \pi r^2 \epsilon,
\end{eqnarray}
where $p$, $\epsilon$ and $m$, are the pressure, the energy density and the gravitational mass respectively, as measured in a proper frame of reference, and $u^{\mu}$ is the fluid's four-velocity. All the quantities have well defined values at the radius $r$ and they will be used as inputs in the numerical integration of the oscillation equations for each phase of the star.

The metric function $\nu(r)$ has the boundary condition
\begin{equation}
\label{BoundaryConditionMetricFunction}
    \nu(r=R)= \ln \bigg( 1-\frac{2M}{R} \bigg),
\end{equation}
where $R$ is the radius of the star and $M\doteq m(R)$ its total mass (as measured by faraway observers). With this condition the  metric function $\nu(r)$ will match smoothly the Schwarzschild metric outside the star and will have a definite value at its origin. Another boundary condition to the system naturally is $m(0)=0$, which expresses the regularity of the metric at the center of the star. Finally, we must have $p(R) =0$ at the stellar surface.

\section{Radial Oscillation Equations}

The first form of the radial oscillation equations can be found in Chandrasekhar's pioneering article \citep{1964PhRvL..12..114C}, in which the  main goal was the study of the dynamical stability of relativistic compact stars. After his work, there was a lot of interest in the understanding of the behavior of compact stars under radial disturbances. Since our objective here is to study phase transition's effects on the dynamical stability  of compact hybrid stars, we will not scrutinize the homogeneous (one-phase) case. For aspects of the stellar stability under radial perturbations without phase transitions, see e.g. \cite{Glendenning-book,haensel2007neutron, Wheeler1965}.

Chandrasekhar's work showed that the equations governing each phase (quark or hadronic) can be obtained by making perturbations in fluid and spacetime variables and inserting them into Einstein's and baryon number conservation equations.  When first order terms are retained a second order differential equation is obtained. For reasons pertaining to advantages in numerical treatments, the oscillation equation can be  split into two first order equations. This has been done by \cite{1992A&A...260..250V}, who derived a set of first order equations for the quantities $\Delta r/r $ and $\Delta p/p$ (see Eqs. (9) and (10) of \cite{1992A&A...260..250V}). More recently, \cite{1997A&A...325..217G} obtained a set
of equations for the relative radial displacement $\Delta r/r $ and the Lagrangian perturbation of
the pressure $\Delta p$ (see Eqs. (11) and (12) of \cite{1997A&A...325..217G}). These sets of equations
are equivalent. In this work we adopt the equations of \cite{1997A&A...325..217G}, because they are particularly suitable for numerical applications and boundary conditions in terms of them have a simpler physical interpretation and a more natural derivation. Another important advantage of this system of
oscillation equations is that they do not involve any derivatives of the adiabatic
index $\Gamma$. Adopting $G = c = 1$ we have

\begin{equation}\label{ecuacionparaXI}
\frac{d\xi}{dr}=V(r)\xi+W(r)\Delta p,
\end{equation}
\begin{eqnarray}\label{ecuacionparaP}
\frac{d\Delta p}{dr}= X(r) \xi + Y(r)  \Delta p,
\end{eqnarray}
with the coefficients given by
\begin{eqnarray}
V(r) &=& -\frac{3}{r}-\frac{dp}{dr}\frac{1}{(p+\epsilon)},  \\
W(r) &=& -\frac{1}{r}\frac{1}{\Gamma p}   \\
X(r) &=& \omega^{2}e^{\lambda-\nu}(p+\epsilon)r-4\frac{dp}{dr} \nonumber \\
 & & + \bigg(\ \frac{dp}{dr}\bigg)^{2}\frac{r}{(p+\epsilon)}-8\pi e^{\lambda}(p+\epsilon) pr    \\
Y(r) &=& \frac{dp}{dr}\frac{1}{(p+\epsilon)}- 4\pi(p+\epsilon)r e^{\lambda} \label{ecuacionparaP},
\end{eqnarray}
where $\omega$ is the eigenfrequency and the quantities $\xi \equiv \Delta r / r$ and $\Delta p$ are assumed to have a harmonic time dependence ($\varpropto e^{i\omega t})$.
To solve equations (\ref{ecuacionparaXI})--(\ref{ecuacionparaP}) in each phase of the star one needs two boundary conditions there.
The  condition of regularity at $r=0$ requires that the coefficient associated with the $1/r$
term in Eq. (\ref{ecuacionparaXI}) must vanish \citep{1992A&A...260..250V,1997A&A...325..217G,1999A&A...344..117G}. Thus, we have      
\begin{equation}\label{DeltaP}
(\Delta p)_{r=0}=-3(\xi \Gamma p)_{r=0}.
\end{equation}
Note that the eigenfunctions can be normalized in order to have $\xi(0)=1$. The surface of the star is determined by the condition that for $r\rightarrow R$, one has $p \rightarrow 0$ at all times. This implies that the Lagrangian perturbation in the pressure at the stellar surface is zero. Therefore another boundary condition is
\begin{equation}\label{PenSuperficie}
(\Delta p)_{r=R}=0.
\end{equation}

Boundary conditions for possible interfaces splitting different phases of a star are still in order, which we turn our attention to in the next section.

\section{Junction conditions at the interface}
\label{junction}

We now discuss the junction conditions needed for the numerical integration of the oscillations' equations when a sharp interface due to a first order phase transition takes place inside a hybrid compact star. Such conditions are intrinsically related to the velocity of the phase transition near the surface splitting any two phases (see \cite{1989A&A...217..137H} for further details). We discuss them in what follows.

\subsection{Slow Transitions}

When the  characteristic timescale of the process transforming one phase into another is much larger than those of the perturbations, we are in the scenario of slow phase transitions. In this case, volume elements near a surface splitting two phases do not change their nature due to the perturbations but they just co-move with the splitting surface, stretching and squashing (due to pressure changes), and there is no mass transfer from one phase to another. This implies that the jump of $\xi$ across the interface, $[\xi]^+_-\equiv \xi^+-\xi^-$, should always be null because one can always track down such near surface elements:
\begin{equation}
[\xi]^+_- = 0 .    
\label{junction_xi_slow}
\end{equation}

The instantaneous constancy of the pressure (different from the equilibrium value, though) on both sides of the splitting-phase surface (directly related to the absence of thin shell surface quantities, such as thin shell surface tensions and energy densities; see, e.g., \cite{2014PhRvD..90l3011P}) also guarantees that 
\begin{equation}
[\Delta p]^+_-=0.    
\label{junction_dp_slow}
\end{equation}
These results can be rigorously derived by making use of generalized distributions \citep{2015ApJ...801...19P}.

In a more general case when the interface has thin shell surface degrees of freedom, the dynamics of radial oscillations could be modified.
As one expects from the lack of transmutation of underneath surface volume elements as regards a phase-splitting surface into above ones (and vice-versa), even when thin shell surface effects take place we still have $[\xi]^+_-=0$. Nevertheless, due to the presence of a nontrivial surface dynamics, $\Delta p$ ceases to be continuous \citep{2015ApJ...801...19P}.  Although interesting, in the present work we limit ourselves to the situations in which thin shell surface degrees of freedom are absent when the system is in equilibrium.

\subsection{Rapid Transitions}
\label{fasttransitions}

Rapid phase transitions are characterized by conversion rates transforming one phase into another whose timescales are much smaller than those of the perturbations \citep{1989A&A...217..137H}. For practical purposes this means instantaneous change of nature of volume elements (from one state of matter to another) near a phase-splitting boundary $H$ due to perturbations. This would imply mass transfers between the two phases. 
Since the conversion rates are very fast, the surface $H$ that splits two any phases can be seen in thermodynamic equilibrium at all times, which means one should characterize it by a constant pressure, equal to that in the absence of perturbations, implying thus $[p]^+_-=0$, which in turn would lead $\Delta p$ to have a null jump across the surface $H$:
\begin{equation}
[\Delta p]^+_-=0.    
\label{junction_dp_rapid}
\end{equation}
Given the practically instantaneous change of nature of volume elements near $H$, one would imagine it would be hard to keep trace of them. Thus, one would expect the Lagrangian displacements of volume elements immediately above and below $H$ to be discontinuous, which would be the key difference between rapid and slow phase transitions regarding boundary conditions. 

We now deduce the boundary condition for $\xi$ in the case of rapid phase transitions by means of physical considerations alone. 
This will be obtained by demanding that $H$ be well-localized, i.e., $[r_H]^+_-=0$, where $r_{H}^{\pm}$ is the radial position of the phase-splitting surface with respect to the radial coordinates above and below it, respectively. Let us consider that in equilibrium $H$ is at the position $r_{H}^{\pm}=R_0$. When perturbations take place, we should generically have $r_{H}^{\pm}=R_0+{\cal A}^{\pm}$, where ${\cal A}^{\pm}$ so far are unknowns and obviously are of the order of $\Delta r\equiv \bar{\xi}$. From the definition of the Lagrangian displacement of the pressure, it follows that
\begin{equation}
p(r,t)=p_0(r)+\Delta p (r,t) - p_0' \bar{\xi}(r,t)\label{p},
\end{equation}
where we have defined the prime operation as the radial derivative and $p_0(r)$ stands for the pressure at $r$ in the absence of perturbations.
At $r=r_H^{\pm}$, due to the nature of rapid phase transitions, it follows that $p(r_H^{\pm},t)=p_0(R_0)$. We have $p_0(r_H^{\pm})\simeq p_0(R_0)+ (p_0')^{\pm} {\cal A}^{\pm}$. Thus, from the above and Eq. (\ref{p}), it follows that
\begin{equation}
{\cal A}^{\pm}=\bar{\xi}^{\pm} -\frac{\Delta p^{\pm}}{(p_0')^{\pm}}\label{surface_displacement}.
\end{equation}
From the condition $[r_{H}]^+_-=0$ we have 
\begin{equation}
\left[ \bar{\xi} - \frac{\Delta p}{\partial_r p_0}\right]^+_-=0\label{jumpDeltar}.
\end{equation}

We point out that the conditions we have obtained for $\Delta p$ and $\Delta r$ are identical to the ones found in \cite{2004CQGra..21.1559K}, though different reasonings have been used.

For numerical purposes, as we have seen previously, it is more convenient to work with $\xi\equiv \Delta r/r= \bar{\xi}/r$. Since $r$ is always continuous across $H$ we can split both sides of Eq.~(\ref{jumpDeltar}) by it, resulting thus in 
\begin{equation}
\left[\xi -\frac{\Delta p}{r p_0'} \right]^+_-=0
\label{jumpxifast}.
\end{equation}

\subsection{Classical limit of rapid phase transition's boundary condition}

It is instructive to calculate the classical limit of Eq.~ (\ref{jumpxifast}) in order to check if it agrees with the one calculated in \cite{1989A&A...217..137H}. This can be easily done by assuming that $2m/r\ll 1$ and $p/\epsilon\ll 1$ from Eq. (\ref{tov1}), which leads to
\begin{equation}
p_0'(r)\approx -\frac{Gm(r)\rho(r)}{r^2}\label{pres_clas},
\end{equation}
where we restored the units only for future convenience and $\rho$ is the mass density of the system. It is also known in the classical case that \citep{1986bhwd.book.....S} 
\begin{equation}
\Delta p=-\frac{\rho v_s^2}{r^2}\frac{d}{dr}(r^3\xi)\label{Delta_p_class},
\end{equation}
with $v_s^2\equiv \partial p/\partial\rho$, the squared adiabatic speed of the sound. 

From the fact that $[\Delta p]^+_-=0$ and Eqs. (\ref{pres_clas}) and (\ref{Delta_p_class}), it follows from Eq. (\ref{jumpxifast}) that
\begin{equation}
[\xi]^+_-=\frac{(\Delta p)_-}{R_0}\left[ \frac{1}{p_0'}\right]^+_-=\frac{(v_s^2)_-}{Gm(R_0)R_0} \left(\frac{\rho_-}{\rho_+}-1 \right)(r^3\xi)'_-.
\end{equation}
One can check that the above equation agrees with the one obtained in \cite{1989A&A...217..137H}, which means that Eq. (\ref{jumpxifast}) is indeed its relativistic generalization.

\section{Numerical procedure}
\label{numerical}

Let us now succinctly describe the numerical procedure that will be used subsequently. The oscillation equations are solved numerically by means of a shooting method. 
At first, for a given central pressure and for each set of parameters of the EOS, we integrated the Tolman-Oppenheimer-Volkoff stellar structure equations  in order to obtain the coefficients of the oscillation equations.
Then, we start at the core with the numerical integration of Eqs. (\ref{ecuacionparaXI})--(\ref{ecuacionparaP}) for a trial value of $\omega^2$ and a given set of values of $\xi$ and $\Delta p$ such that the boundary condition at the center is fulfilled. The equations are integrated outwards until the quark-hadron interface is reached, where we use the junction conditions ({Eqs. (\ref{junction_xi_slow}) and (\ref{junction_dp_slow}) for the slow case and Eqs. (\ref{junction_dp_rapid}) and (\ref{jumpxifast}) for the rapid case}) to obtain the correct values of $\xi$ and $\Delta p$ at the other side of the interface. Thereafter, the integration proceeds outwards trying to match the boundary condition at the stellar surface. After each integration, the trial value of $\omega^2$ is corrected until the desired precision on the boundary condition is achieved. The discrete values of $\omega$ for which the oscillations equations are satisfied are the eigenfrequencies of the star (for more details see \cite{2010PhRvD..82f3006V}). 

We have also implemented a shooting to a fitting point method in which two ``shots'' are made, one from the center and the other from the surface of the star,  trying to match the junction conditions at the quark-hadron interface. Both methods produced identical results.

\begin{figure}[tb]
    \centering
    \includegraphics[width=\columnwidth]{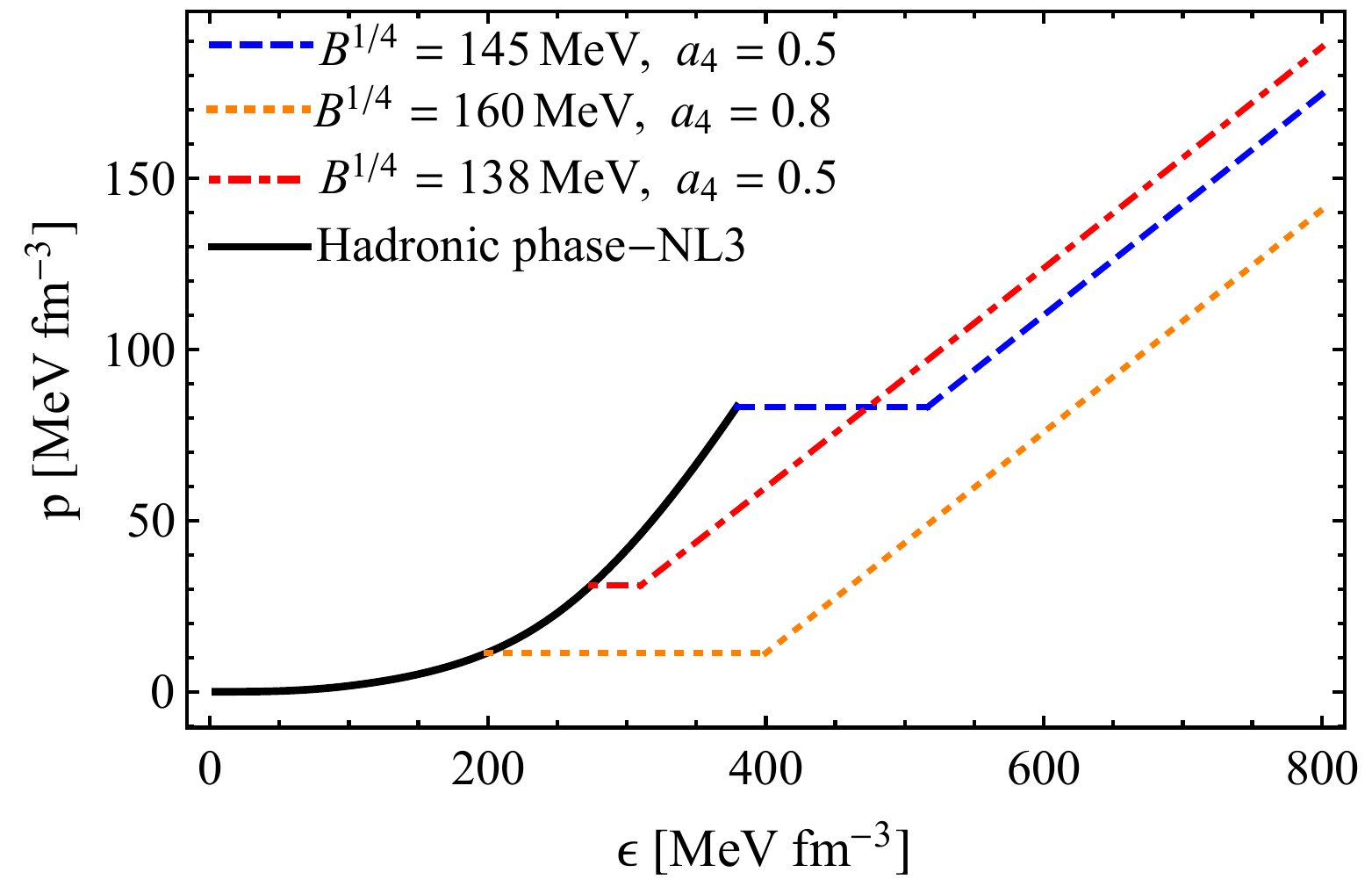}
   \caption{Some equations of state for hybrid stars using the hadronic and quark models of Sec. \ref{section_eos}. We adopt $a_2=(100$~MeV)$^2$ and the parameters $B$ and $a_4$ fulfill Eq. (\ref{B_min_hyb_stars}). For each set of EOS parameters, the sharp interface between the hadronic and the quark phase is determined by requiring the equality of the pressure and the Gibbs free energy per baryon in both phases.} 
\label{eos}
\end{figure}

\begin{figure}[tb]
    \centering
    \includegraphics[width=\columnwidth]{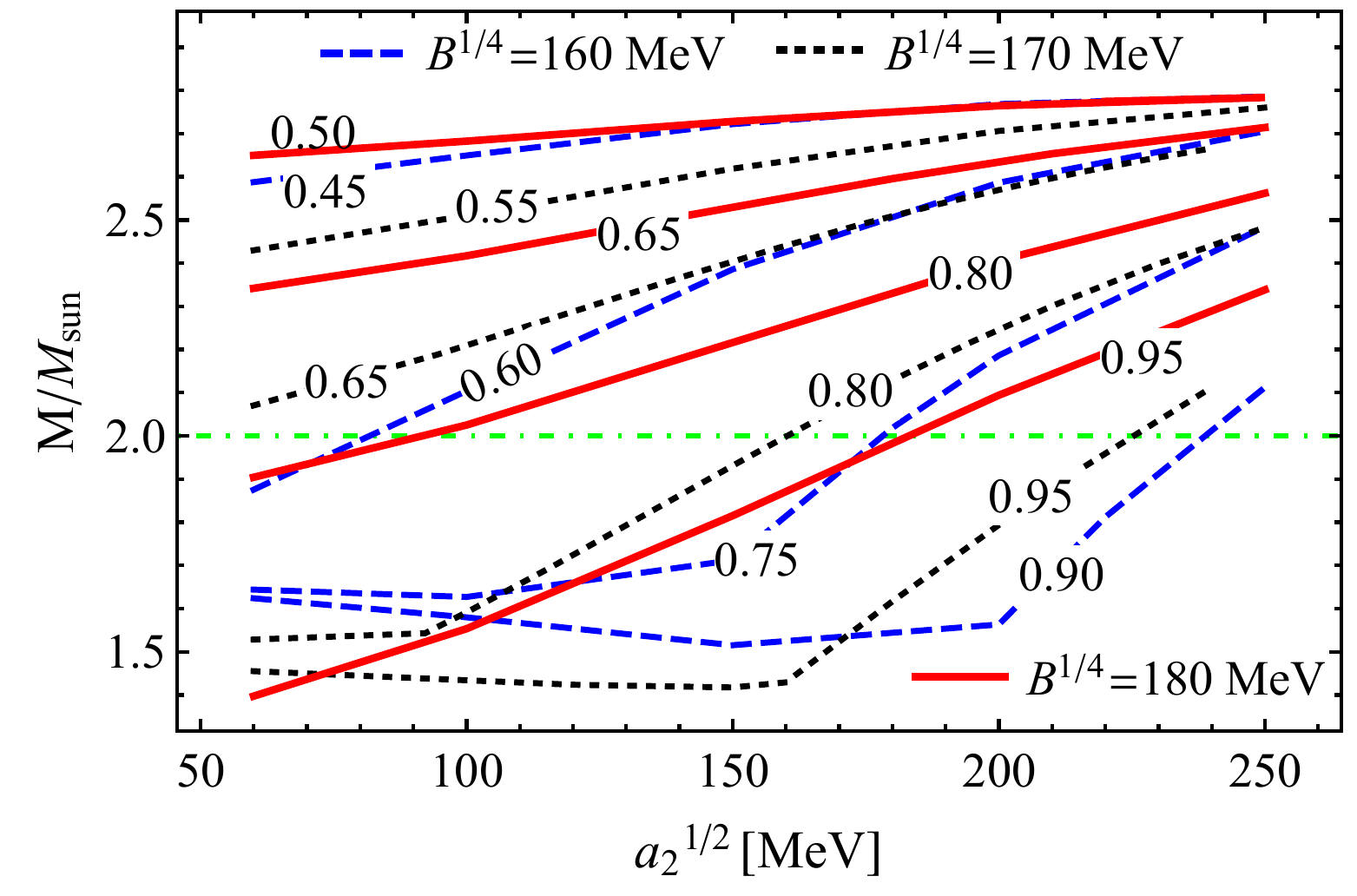}
    \caption{Maximum masses of hybrid stars in hydrostatic equilibrium for some selected values of $B$, $a_4$ (numbers on the curves) and $a_2$. There are several configurations which lead to masses larger than $2M_{\odot}$. For a given $a_2$ and $B$, one sees that the smaller the $a_4$ the larger is the maximum mass of the hybrid star. For a given $a_4$ and $B$, in most cases the maximum mass increases with $a_2$. }
\label{mmax}
\end{figure}

\begin{figure*}[tbh]
    \centering
    \includegraphics[width=\columnwidth]{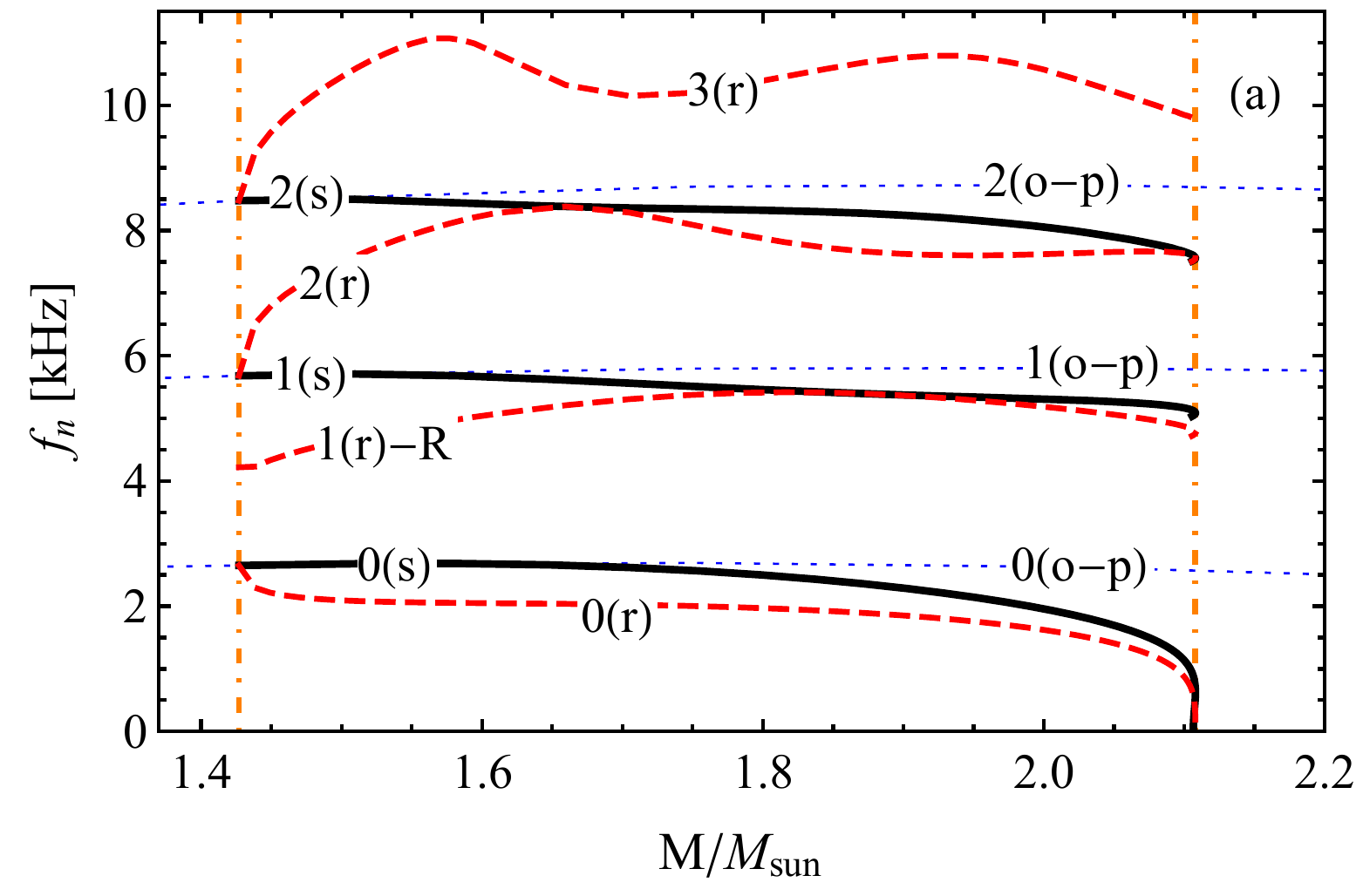}
    \includegraphics[width=\columnwidth]{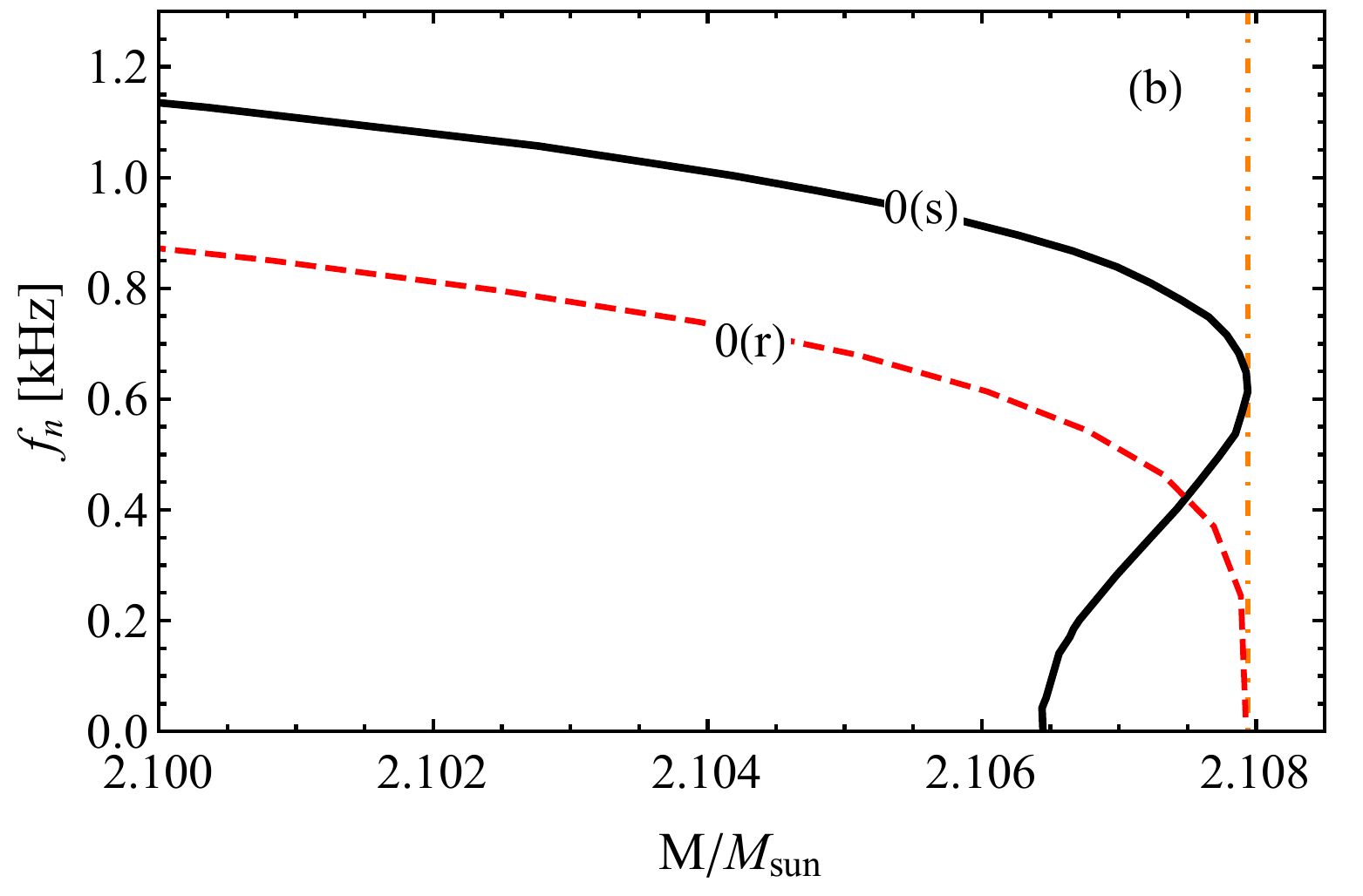}
    \caption{(a) Oscillation modes for rapid (r), slow (s) and one-phase (o-p) stars for the case $B^{1/4}= 138$ MeV, $a_4=0.5$ and $a_2^{1/2}=100$ MeV.  The numbers in front of the letters stand for the associated modes considered, with $0$ the fundamental one. Note that 1(r)-R, the reaction mode, is exclusively related to rapid phase transitions and its frequency does not converge to any eigenfrequency associated with slow phase transitions and one-phase stars around the mass hybrid stars start to exist ($\sim 1.4 M_{\odot}$, as indicated by the vertical dot-dashed curve). Modes associated with different kinds of phase transitions could intersect for certain masses.  (b) Fundamental modes near the maximum mass $M_{max}\approx 2.11$$M_{\odot}$. For rapid phase transitions the frequency goes to zero as the mass approaches $M_{max}$. However, for slow phase transitions $M_{max}$ is reached before the frequency of the fundamental mode vanishes; i.e. some stellar configurations beyond $M_{max}$ (higher central densities) are stable under small radial perturbations.}
\label{B_138_a4_0_5}
\label{B_138_a4_0_5_max_mass}
\end{figure*}

\section{Results}
\label{results}

When talking about the stability of hybrid systems, independent of whether slow or rapid phase transitions take place, the fundamental mode with frequency $f_0$ is crucial. It is defined as the nodeless solution to Eqs.~(\ref{ecuacionparaXI})--(\ref{ecuacionparaP}), supplemented by Eqs. (\ref{tov1}), (\ref{tov2}) and (\ref{tov3}). When it is purely imaginary, the star is unstable because perturbations could grow without limits. We have previously seen that hybrid stars only exist for certain values of the effective bag constant larger than a minimum one, for given $a_2$ and $a_4$, defined by Eq. (\ref{B_min_hyb_stars}) (see Fig. \ref{hyb_star_stable_QS}). In our forthcoming analysis this fact is taken into account. Besides, for the hadronic phase of the star, we make use of the relativistic mean field EOS with the NL3 parametrization already presented in Sec. \ref{section_eos}. We assume that the phase transition is sharp (no mixed phase), and determine the interface by imposing the Gibbs equilibrium conditions (equality of the pressure and the Gibbs free energy per baryon of the quark and hadronic phases). Figure \ref{eos} shows some examples of equations of state we will work with.

In Fig. \ref{mmax} we calculate the maximum masses associated with some $B$, $a_4$ and $a_2$. As clearly seen from this figure, several different configurations result in maximum masses larger than $2M_{\odot}$, in agreement with the recent discovery of two very high mass pulsars  \citep{Demorest2010,Antoniadis2013}. Naturally, these are of main physical interest but other less massive configurations are also interesting for having insights into the macroscopic fingerprints and conspicuous aspects of different phase transitions. In our analysis we take into consideration both above-mentioned characteristics.

\subsection{Rapid versus slow modes}

In Fig. \ref{B_138_a4_0_5} we calculate the fundamental and some excited modes assuming rapid and slow transitions for a sequence of stars with the quark EOS parametrization $B^{1/4}= 138$ MeV, $a_4=0.5$ and $a_2^{1/2}=100$ MeV. Hybrid stars in this case have maximum mass around $2.11 M_{\odot}$ and the pressure where the phase transition occurs is around 31 MeV fm$^{-3}$.   Notice that the fundamental modes of rapid and slow phase transitions tend to the same value in the limit the system becomes one-phase (purely hadronic, occurring when the radius of the innermost phase goes to zero). 
Additionally, for $m\geq 2$, the $m$th excited rapid mode matches continuously the $(m-1)$th slow one when the quark core shrinks. 

Note that some rapid transition frequencies could overlap slow transition ones. The fundamental frequencies associated with rapid phase transitions are smaller than their slow counterparts for all possible masses of the system. As indeed expected from the densities involved, frequencies for the modes are ordinarily around some kHz. We do not elaborate on aspects of one-phase stars because they are already very well known \citep{Glendenning-book,haensel2007neutron, Wheeler1965}.

\subsection{The reaction mode}

Regarding rapid phase transitions, a new mode appears, which is the general relativistic generalization of the reaction mode discovered by \cite{1989A&A...217..137H}. We find that for hybrid neutron stars with some solar masses the reaction mode would be either be the first excited or the fundamental one, depending upon the EOS parametrizations. For example, if the NL3 parametrization of the hadronic equation of state is used and we take $a_2=(100$~MeV)$^2$, then the reaction mode is the fundamental mode when $B^{1/4}\gtrsim 140$ MeV.
Indeed, as one can see for the case $B^{1/4}=138$ MeV and $a_4=0.5$ of Fig. \ref{B_138_a4_0_5}, the reaction mode is the first excited eigenfrequency for each configuration. It does not converge to any slow or one-phase mode when the radius of the core goes to zero, being thus the fingerprint of rapid phase transitions. The  main relevance of the reaction mode to the stability of the star is naturally when it is the fundamental mode. 

\begin{figure}[tb]
    \centering
    \includegraphics[width=\columnwidth]{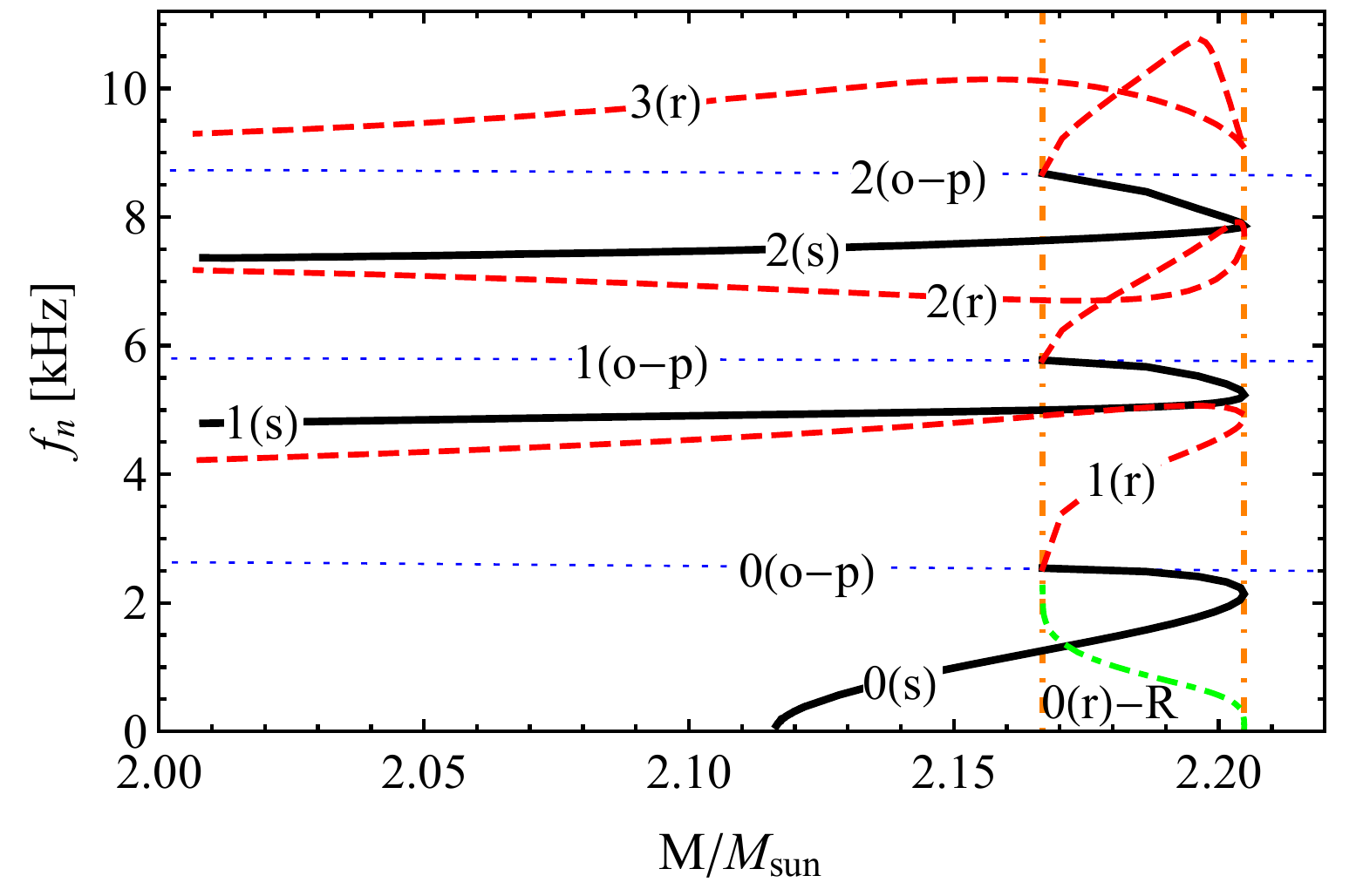}
    \caption{First radial modes for the case $B^{1/4}= 145$ MeV, $a_4=0.5$ and $a_2^{1/2}=100$ MeV. The phase transition pressure in this case is approximately $83 \, \mathrm{MeV} \, \mathrm{fm}^{-3}$ and the maximum mass of the system is $2.21$$M_{\odot}$. One can clearly notice that the reaction mode is now the fundamental mode for rapid phase transitions. Besides, it exists right from the onset of hybrid stars, which occurs for the mass $2.17$$M_{\odot}$. We point out that $f_{0(r)-R}$ is different from the frequency of the modes $1$(r), $0$(o-p) and $0$(s) at such mass.   
Additionally, $f_{0(r)-R}$ vanishes for the maximum mass model and therefore the region where the frequency $f_{0(r)-R}$ exists as a real number is intrinsically associated with $\partial M/\partial \rho_c\geq 0$, as easily seen from Fig. \ref{B_138_a4_0_5_mass_rhoc}.} 
\label{B_145_a4_0_5}
\end{figure}

Figure \ref{B_145_a4_0_5} exemplifies a situation contrasting with Fig. \ref{B_138_a4_0_5} where $B^{1/4}$ is increased beyond $140$ MeV and the reaction mode is now the fundamental one.
As one further increases $B$ for fixed $a_4$ and $a_2$, a maximum value is reached above which the frequency of the fundamental mode of rapid phase transitions becomes purely imaginary. We will come back to this issue later when we give their estimates and associated star aspects for some selected $a_4$. 

\begin{figure}[tb]
    \centering
    \includegraphics[width=\columnwidth]{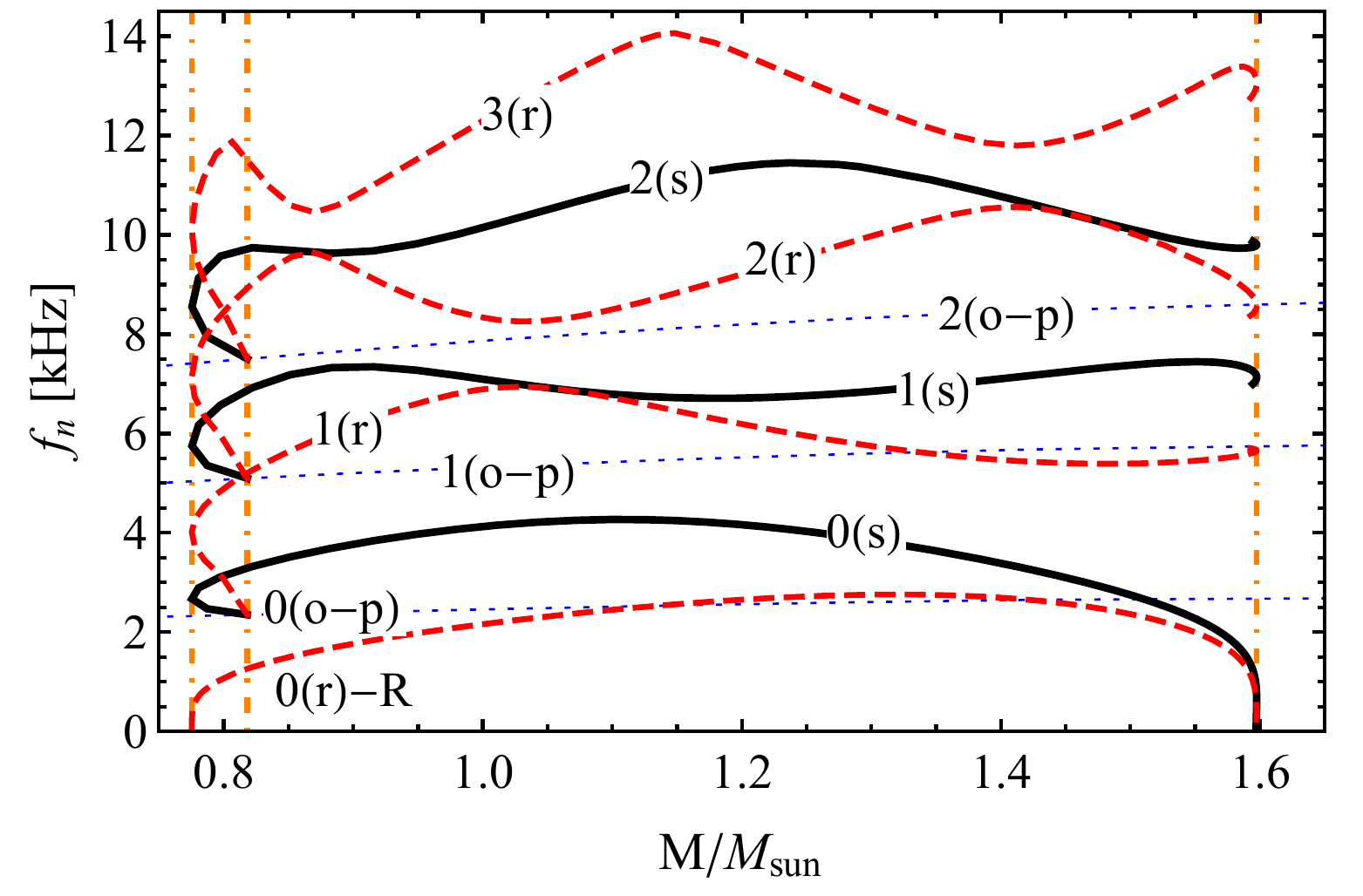}
    \caption{First radial modes for the case $B^{1/4}= 160$ MeV, $a_4=0.8$ and $a_2^{1/2}=100$ MeV. The maximum mass for these parameters is around $1.6 M_{\odot}$ and the transitional (or plateau) mass (from a one-phase to a hybrid star) is $M_c\approx 0.82 M_{\odot}$ (indicated by the second vertical dot-dashed line). The numbers and letters on the curves have the same meaning as in Fig. \ref{B_138_a4_0_5}. Here the reaction mode corresponds to the fundamental eigenfrequency of rapid phase transitions and it does not exist for very small quark cores, differently from the case of slow phase transitions.} 
\label{B_160_a4_0_8}
\end{figure}

Since when one increases $a_4$ for a fixed $a_2$ the minimum value of $B$ such that hybrid stars exist also increases (see Fig.~\ref{hyb_star_stable_QS}), the reaction mode will eventually only be the fundamental one.
This is exemplified by Fig. \ref{B_160_a4_0_8} for the case $B^{1/4}=160$ MeV and $a_4=0.8$. Observe in this case that there is a region around the critical mass marking the onset of hybrid stars where the reaction eigenfrequencies do not exist, while they do for slow phase transitions. We discuss this issue in the next subsection.

One could also calculate the reaction mode when the quark core is small. It is classically known \citep{1989A&A...217..137H} that in this case it is related to the jump of densities in the phase transitions, $\eta\doteq \epsilon_-/\epsilon_+$, where $\epsilon_-$ is the energy density at the top of the quark core and $\epsilon_+$ is the energy density at the base of the hadronic phase. More specifically, $\omega_R^2 \sim (3-2\eta)/(\eta-1)$ \citep{1989A&A...217..137H}. Hence, the natural general relativistic generalization to this classical expression would be $\omega_R^2 \sim (3[1+p_t/\epsilon_+]-2\eta)/(\eta-1)$, where $p_t$ is the phase transition pressure \citep{1971SvA....15..347S,1987A&A...172...95Z}.
Indeed, we have numerically checked that the relationship is linear. For the case $a_2^{1/2}=100$ MeV, we found that $\omega_R^2/10^8\approx 0.76\,(3-2\eta +3p_t/\rho_+)/(\eta -1)$.
Other values of $a_2^{1/2}$ are related to (slightly) different slopes of $\omega_R^2/10^8$ as a function of $(3-2\eta +3p_t/\rho_+)/(\eta -1)$. Therefore, in principle the reaction mode can be any excited mode, even a very large one if $\eta \rightarrow 1$. However, we note that in our case $\eta$ is not a free parameter because it is unequivocally determined by Gibbs equilibrium condition and the EOSs used. For the cases where the background masses of the neutron stars are of some solar masses, we have not found situations where $\eta \rightarrow 1$ and this explains why the reaction mode was not found to be an overtone. Situations where the reaction mode does not exist is guaranteed if $(3-2\eta +3p_t/\rho_+)<0$, satisfied for large enough $\eta$. Indeed, this inequality is verified to the case related to Fig.~\ref{B_160_a4_0_8}.

\begin{figure}[tb]
    \centering
    \includegraphics[width=\columnwidth]{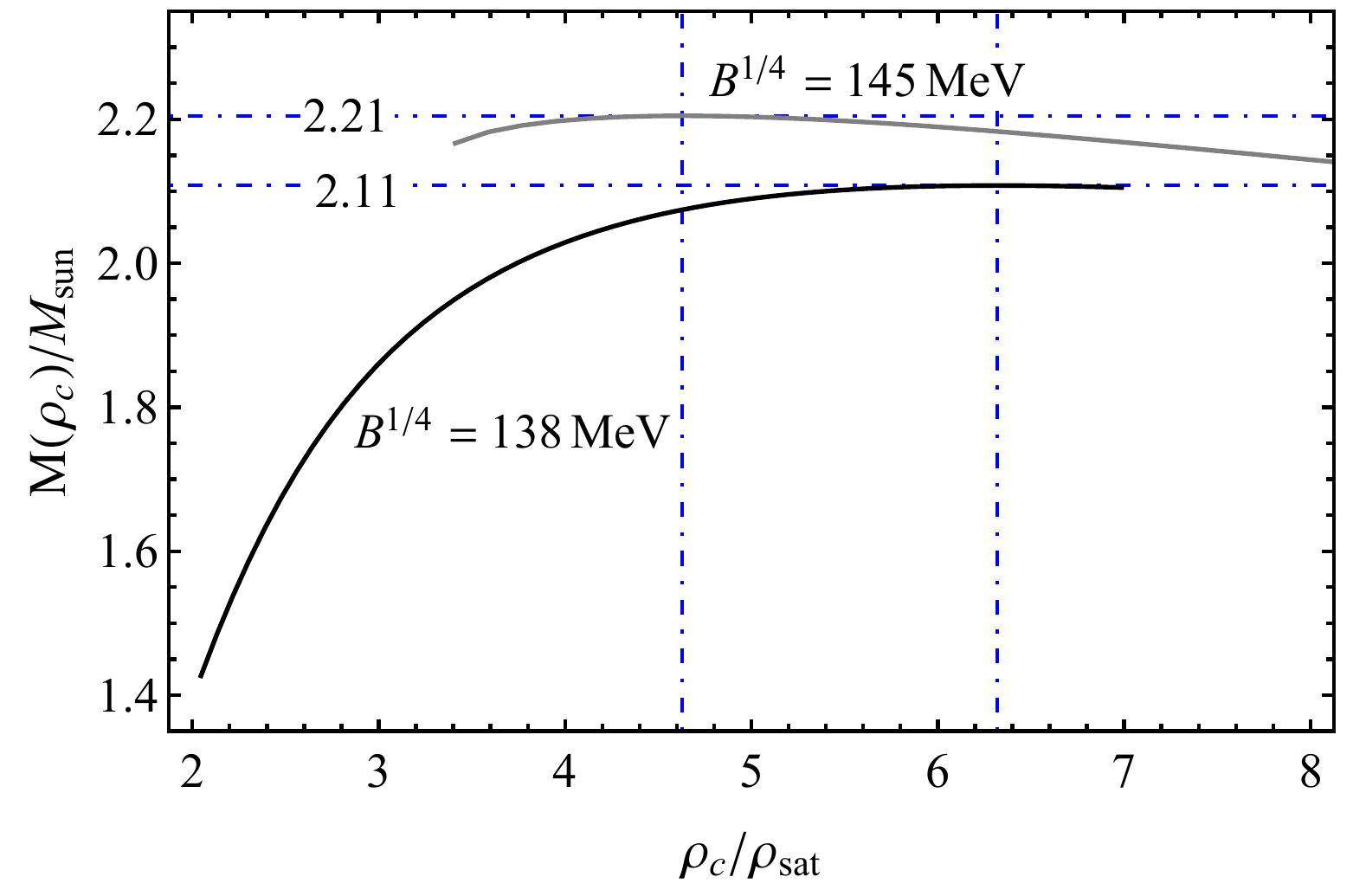}
    \caption{Mass as a function of the central density (in units of the nuclear saturation density $\rho_{sat}= 2.7\times 10^{14}$ g cm$^{-3}$) for a family of hybrid stars with parameters $B^{1/4}=138$ MeV (black solid curve) and $B^{1/4}=145$ MeV (gray solid curve) for $a_4=0.5$ and $a_2^{1/2} =100$ MeV. 
In the case of rapid phase transitions, the conditions $\partial M/\partial \rho_c=0$ and $f_{0(r)}=0$ are fulfilled for the same value of the stellar mass, which turns out to be the maximum mass (c.f. Figs. \ref{B_138_a4_0_5_max_mass} and \ref{B_145_a4_0_5}). Therefore, models beyond the maximum mass are unstable under radial perturbations. 
In the case of slow phase transitions the condition $f_{0(s)}=0$  is fulfilled for a mass beyond the maximum mass (at a point of the curve where $\partial M/\partial \rho_c<0$): for the black thick curve this happens for a mass very close to $2.11$$M_{\odot}$, while for the gray curve the associated slow zero-frequency mass is $2.12$$M_{\odot}$ (see Figs. \ref{B_138_a4_0_5_max_mass} and \ref{B_145_a4_0_5} for further details). 
This evidences the stability of some stars with central densities larger that the central density of the maximum mass star.} 
\label{B_138_a4_0_5_mass_rhoc}
\end{figure}

\begin{figure}[tb]
    \centering
    \includegraphics[width=\columnwidth]{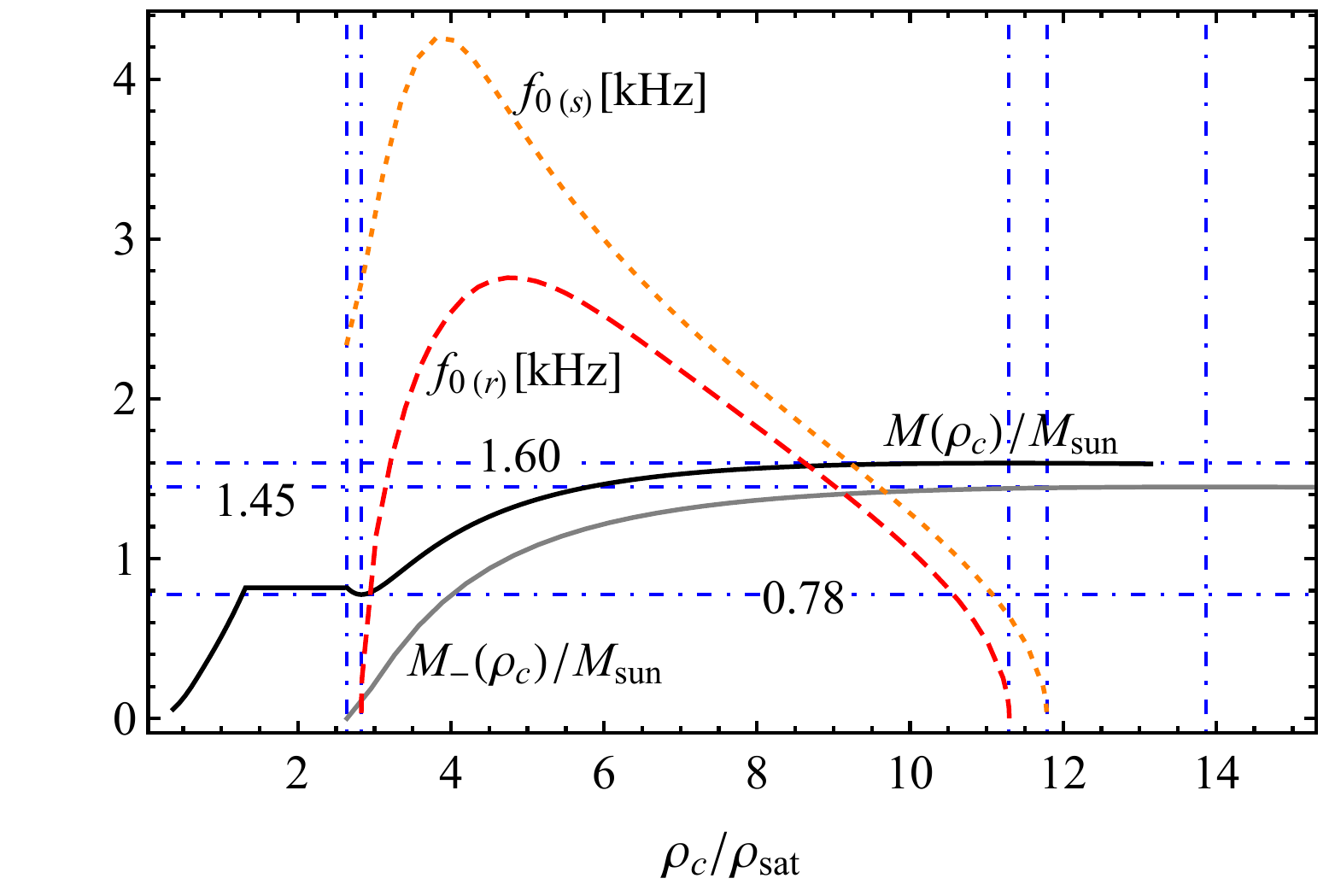}
    \caption{Masses and fundamental eigenfrequencies of a sequence of stars with different central densities with the same parameters of Fig. \ref{B_160_a4_0_8}. For this case, the mass the system changes from a one-phase configuration to a hybrid one is approximately $0.82 M_{\odot}$, represented by the flat part of the black solid curve. In the hybrid sector, there is a set of densities for which the stars' masses decrease, reaching the minimum $0.78 M_{\odot}$ at $\rho_c/\rho_{sat}\approx 3$. Although not evident, the total mass curve presents another critical point at a larger density, associated with the maximum mass $1.60 M_{\odot}$ ($\rho_c/\rho_{sat}\approx 11.3$). Real fundamental eigenfrequencies of rapid phase transitions $f_{0 (r)}$ only exist when $\partial M/\partial \rho_c\geq 0$, as exactly dictated by the classical stability condition for one-phase systems. This clearly contrasts with their slow counterparts, $f_{0 (s)}$, which also admit $\partial M/\partial \rho_c< 0$. Note that zero frequencies are not related to the critical points of $M_-(\rho_c)$, which in this case is at $\rho_c/\rho_{sat}\approx 14$ ($M_-/M_{\odot} \approx 1.45$).} 
\label{mass_densisty_B_160_a4_0_8}
\end{figure}

\begin{figure}[tb]
    \centering
    \includegraphics[width=\columnwidth]{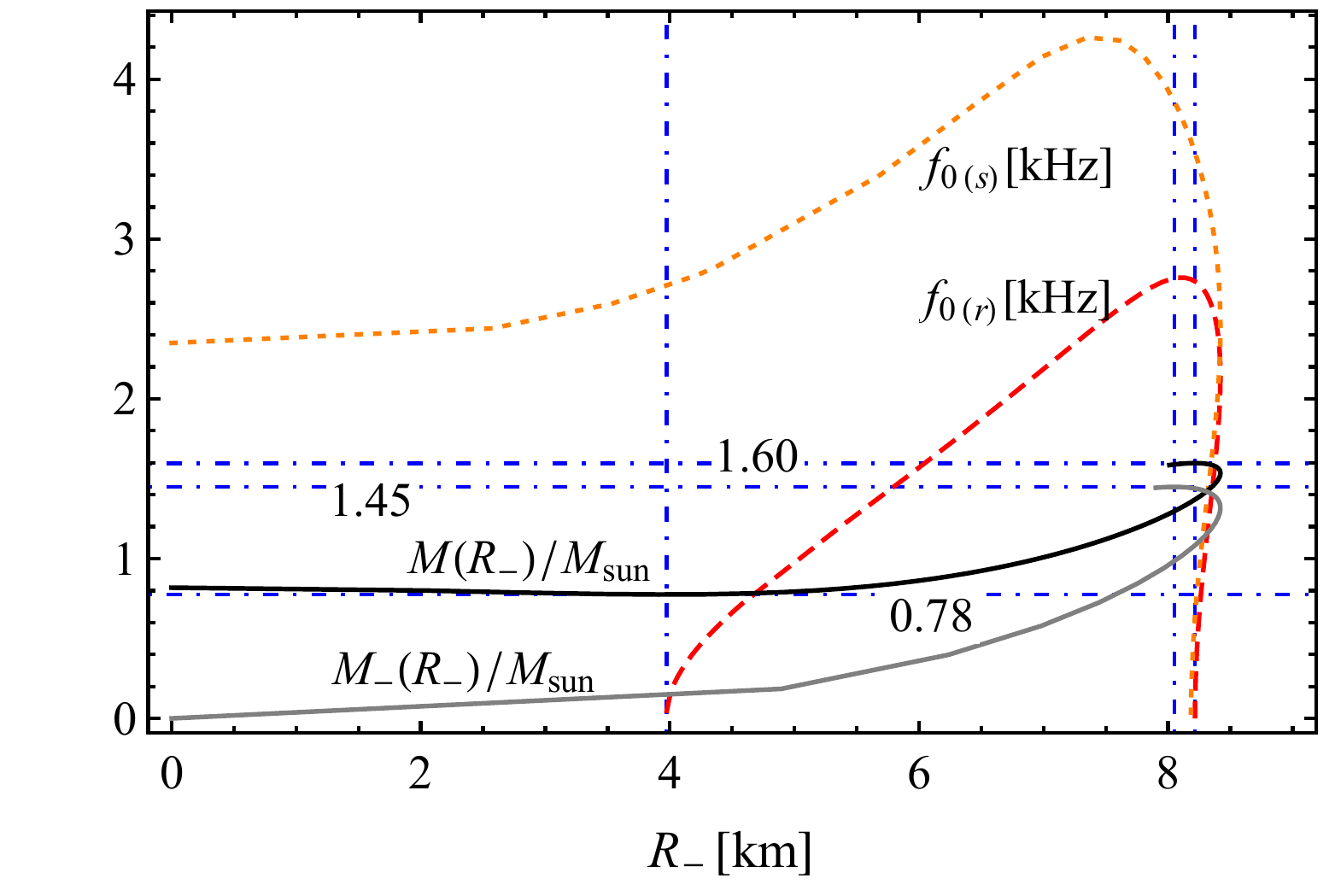}
   \caption{Masses and fundamental eigenmodes of a sequence of stars as a function of their core radii $R_-$ for the same parameters of Fig. \ref{B_160_a4_0_8}. Here one confirms the conclusions of Fig. \ref{mass_densisty_B_160_a4_0_8}: the fundamental rapid mode only exists between the critical points of $M(R_-)$, related to the masses $0.78$$M_{\odot}$ and $1.60 M_{\odot}$, while this is not the case for fundamental slow phase transitions.} 
\label{mass_radius_B_160_a4_0_8}
\end{figure}
%

\begin{figure}[tb]
    \centering
    \includegraphics[width=\columnwidth]{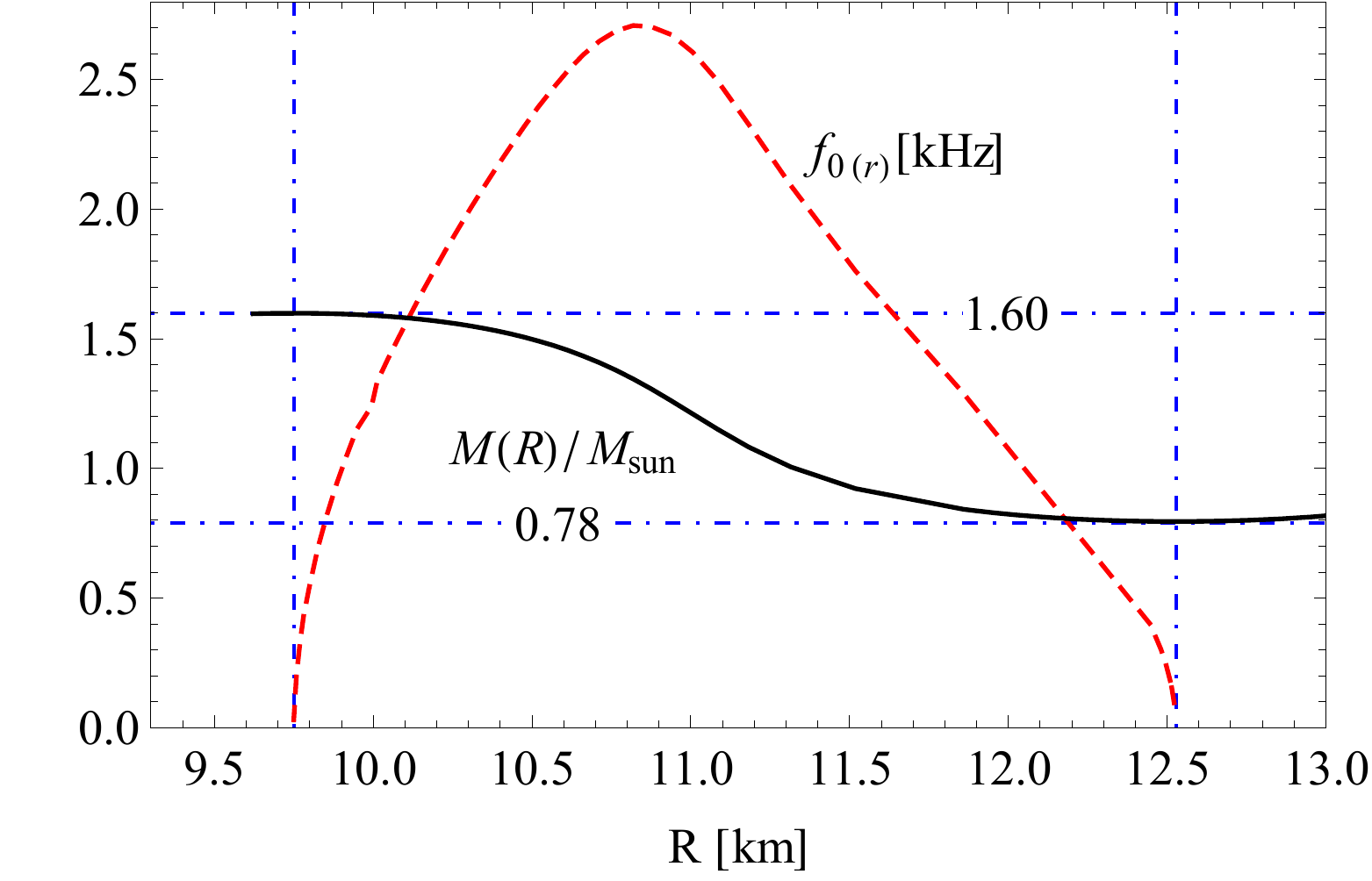}
    \caption{Masses and fundamental rapid eigenmodes of a sequence of hybrid stars as a function of their radii $R$ for the same parameters as in Fig. \ref{B_160_a4_0_8}. The range of radii associated with the existence of the fundamental mode is intrinsically associated with their critical points, which are related to the same masses as in Fig. \ref{mass_radius_B_160_a4_0_8}.} 
\label{mass_radius_ext_B_160_a4_0_8}
\end{figure}
%

\begin{figure}[tb]
    \centering
    \includegraphics[width=\columnwidth]{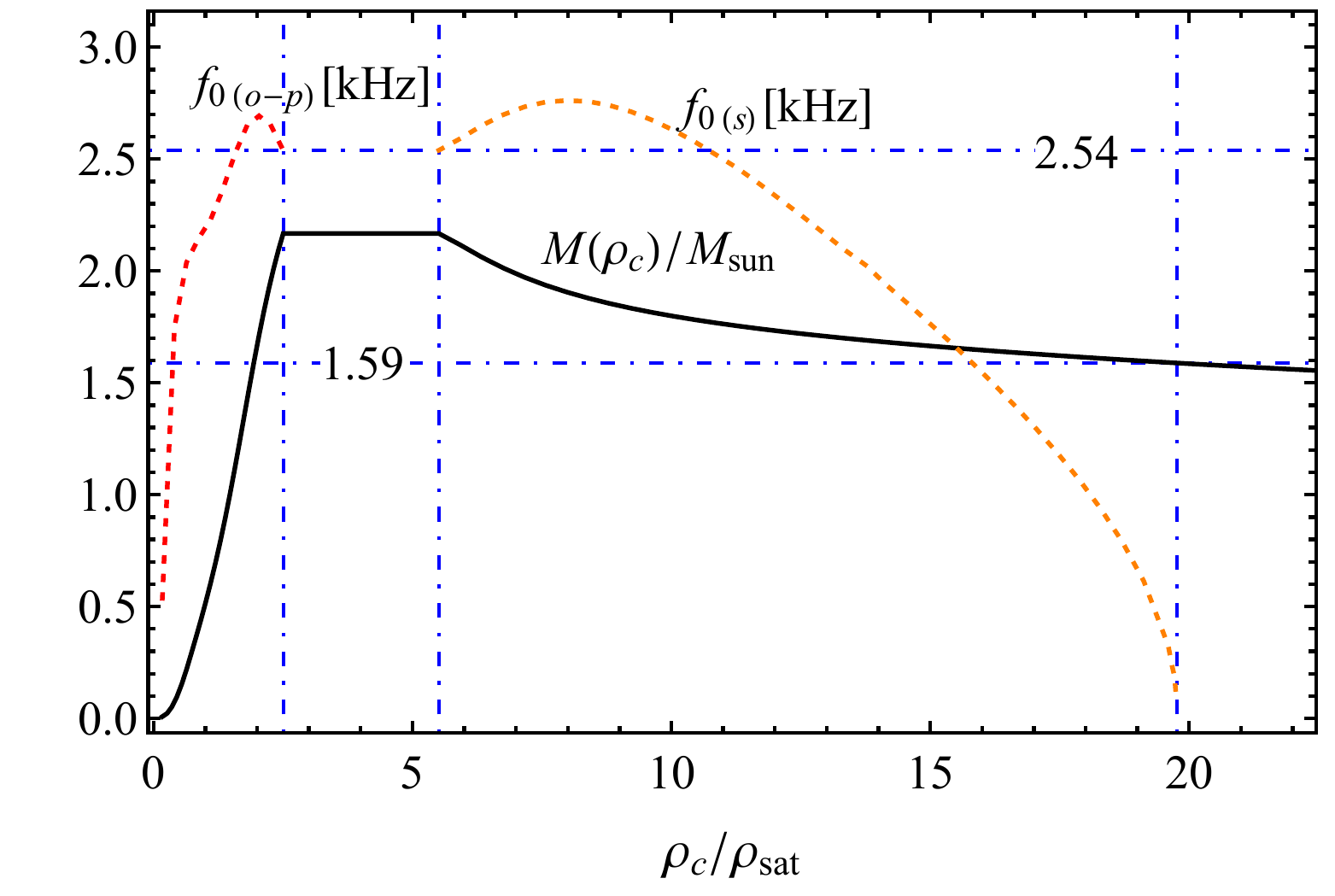}
    \caption{Masses and frequencies (slow hybrid and one-phase) for stars with parameters $B^{1/4}=180$ MeV, $a_4=0.75$ and $a_2^{1/2}=100$ MeV. Note that rapid fundamental frequencies are nonexistent in this case while slow fundamental ones exist up to densities around $5.4\times 10^{15}$ g \,cm$^{-3}$, showing the blatant difference between phase conversions in this case. The mass where the fundamental slow eigenfrequencies is  null for these parameters is $1.59$ M$_{\odot}$.} 
\label{m180}
\end{figure}

\begin{figure}[tb]
    \centering
    \includegraphics[width=\columnwidth]{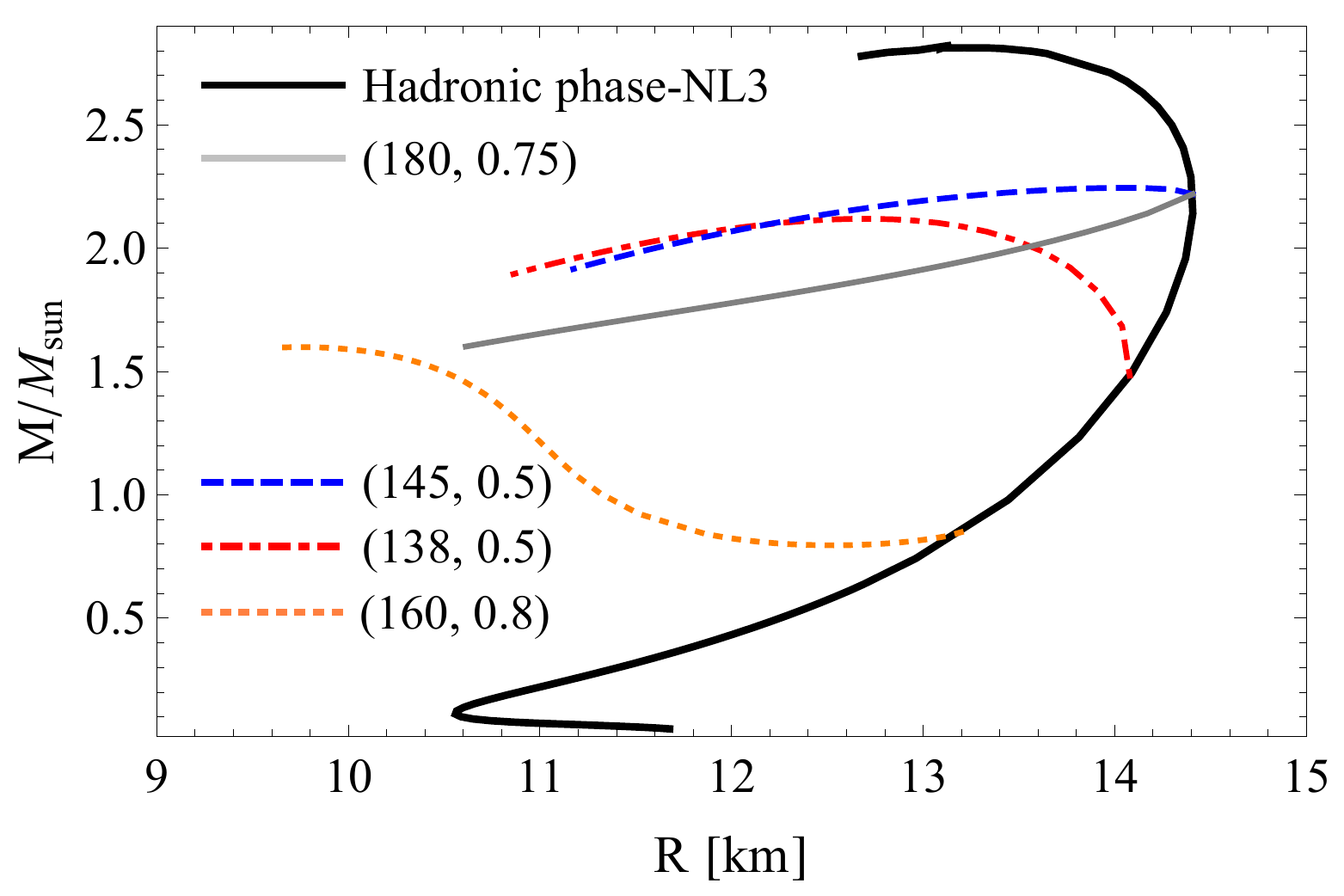}
   \caption{Mass-radius relations for hybrid stars for the models we have investigated. For the legends the first and second terms of each pair corresponds to $B^{1/4}$ and $a_4$, respectively. For all hybrid curves $a_2^{1/2}= 100$ MeV.} 
\label{M-R-relation}
\end{figure}

\subsection{Zero-frequency mass, maximum mass and stellar stability}    

As depicted in Fig. \ref{B_138_a4_0_5_max_mass}, the stellar background mass at which the frequency of the fundamental mode is zero, $M_0$, depends on the nature of the phase transition. The difference between $M_{0,\mathrm{rapid}}$ and $M_{0,\mathrm{slow}}$ for the parameters of Fig. \ref{B_138_a4_0_5_max_mass} is very small, around 0.05\%, while for the case of Fig. \ref{B_145_a4_0_5} it is conspicuous.
Taking the same parameters of Figs. \ref{B_138_a4_0_5_max_mass} and \ref{B_145_a4_0_5}, Fig. \ref{B_138_a4_0_5_mass_rhoc} shows that $M_{0,\mathrm{rapid}}$  coincide with the maximum masses of their $(M,\rho_c)$ plots, and that the reality of the fundamental mode is intrinsically associated with the condition $\partial M/\partial \rho_c\geq 0$. However, for slow phase transitions the zero-frequency mass $M_{0,\mathrm{slow}}$ corresponds to a model with a central density $\rho_c \geq \rho_{c,max}$, being $\rho_{c,max}$ the central density of the model with $M = M_{max}$.
We argue now by means of other examples that this is a generic aspect of rapid and slow phase transitions in hybrid stars and hence a potential way to differentiate them. 

Consider for instance the case associated with Fig. \ref{B_160_a4_0_8}. As clear from Fig. \ref{mass_densisty_B_160_a4_0_8} for rapid phase transitions, differently from slow phase transitions, the system is unstable (nonexistent/imaginary fundamental modes) for the situations where $\partial M/\partial \rho_c<0$. Figs. \ref{mass_radius_B_160_a4_0_8} and \ref{mass_radius_ext_B_160_a4_0_8} confirm that the stability criterion for rapid phase transitions can also be associated with the critical points of the mass as a function of both the quark phase's radius and star's radius, $R_-$ and $R$, respectively. Note, however, that this is not the case when one works with the mass of the inner (quark) phase, $M_-$, as a function of $R_-$ or $\rho_c$ (see Figs. \ref{mass_radius_B_160_a4_0_8} and \ref{mass_densisty_B_160_a4_0_8}, respectively).

Even when $\partial M/\partial \rho_c<0$, stars can also be stable when only slow phase transitions take place. Figure \ref{m180} exemplifies that for a case where the maximum mass of the system exceeds $2M_{\odot}$. 
One clearly sees that the violation range of the condition $\partial M/\partial \rho_c >0$ is considerable: $f_{0(s)}$ is positive up to around four times the central density related to the maximum mass. 
Thus, the usual condition for stability, $\partial M / \partial \rho_c \geq 0$, should in general be superseded because it is dependent on the timescale of the phase conversion.  
Finally, for ease of contrast with other models and completeness, Fig. \ref{M-R-relation} gives the mass-radius relation of all hybrid models investigated previously.

One may wonder whether the above analysis remains valid if one considers that NSs may rotate. In fact, for rotating compact stars the existence of a stable configuration can be threatened  by  nonaxisymmetric  dynamical and secular instabilities.

The most studied  type of rotational dynamical instability is the so called  bar-mode instability. The classical $m=2$ bar-mode instability is excited in Newtonian stars when the ratio $\beta=T/|W|$ of the rotational kinetic energy $T$ to the gravitational binding energy $|W|$ is larger than $\beta =0.27$, but general relativistic effects lower the critical value to $\beta \sim 0.24$ \citep{SBS2000,Saijo01}. 
Differential rotation may change the scenario significantly.  NSs with a high degree of differential rotation may be dynamically unstable for  $\beta \gtrsim 0.01$ \citep{Shibata2002,Shibata2003}.  Additionally, an $m=1$ one-armed spiral instability may become unstable if differential rotation is sufficiently strong \citep{Centrella2001,SBM2003}.  
However,  differential rotation  affects only very young protoneutron stars. In fact, a few minutes after a NS is born in a core collapse supernova, rigid rotation sets in due to the presence of viscosity or a sufficiently strong magnetic field \citep{Haensel2016}. 
Thus, these dynamical instabilities are not expected to have an impact in most cold catalyzed NSs.

Another class of nonaxisymmetric instabilities are  secular instabilities,  which  require the presence of dissipation  due, for example, to viscosity and/or gravitational radiation. 
A particularly interesting class of oscillation modes are $r-$modes, which are large-scale currents in NSs that couple to gravitational radiation and remove energy and angular momentum from the star in the form of gravitational waves \citep{Andersson98,FriedmanMorsink,Friedman78,LOM98}. In the absence of viscous dissipation, they are unstable at all rotation frequencies  leading to an exponential rise of the $r-$mode amplitude. 
However, when viscous damping is taken into account the star is stable at low frequencies but there may remain  instability regions at high frequencies \citep{LOM98,Alford2012b}. 
In addition, the fact that fast-spinning compact stars are observed suggests that nonlinear damping mechanisms are present and limit the exponential grow of $r-$modes that would destroy the NS \citep{Alford2012a}. 
If this instability is stopped at a large amplitude, $r-$modes may be a strong and continuous source of gravitational waves in some objects \citep{Haensel2016}. Another, secular bar-mode instability, sets in at $\beta \simeq  0.14$ \citep{AnderssonComer2017}.

In summary, it is reasonable to expect that the here-presented stability analysis based on radial oscillations should be valid for hybrid stars rotating below a yet unknown but not too small critical frequency.

\begin{figure*}[tb]
    \centering
    \includegraphics[width=\columnwidth]{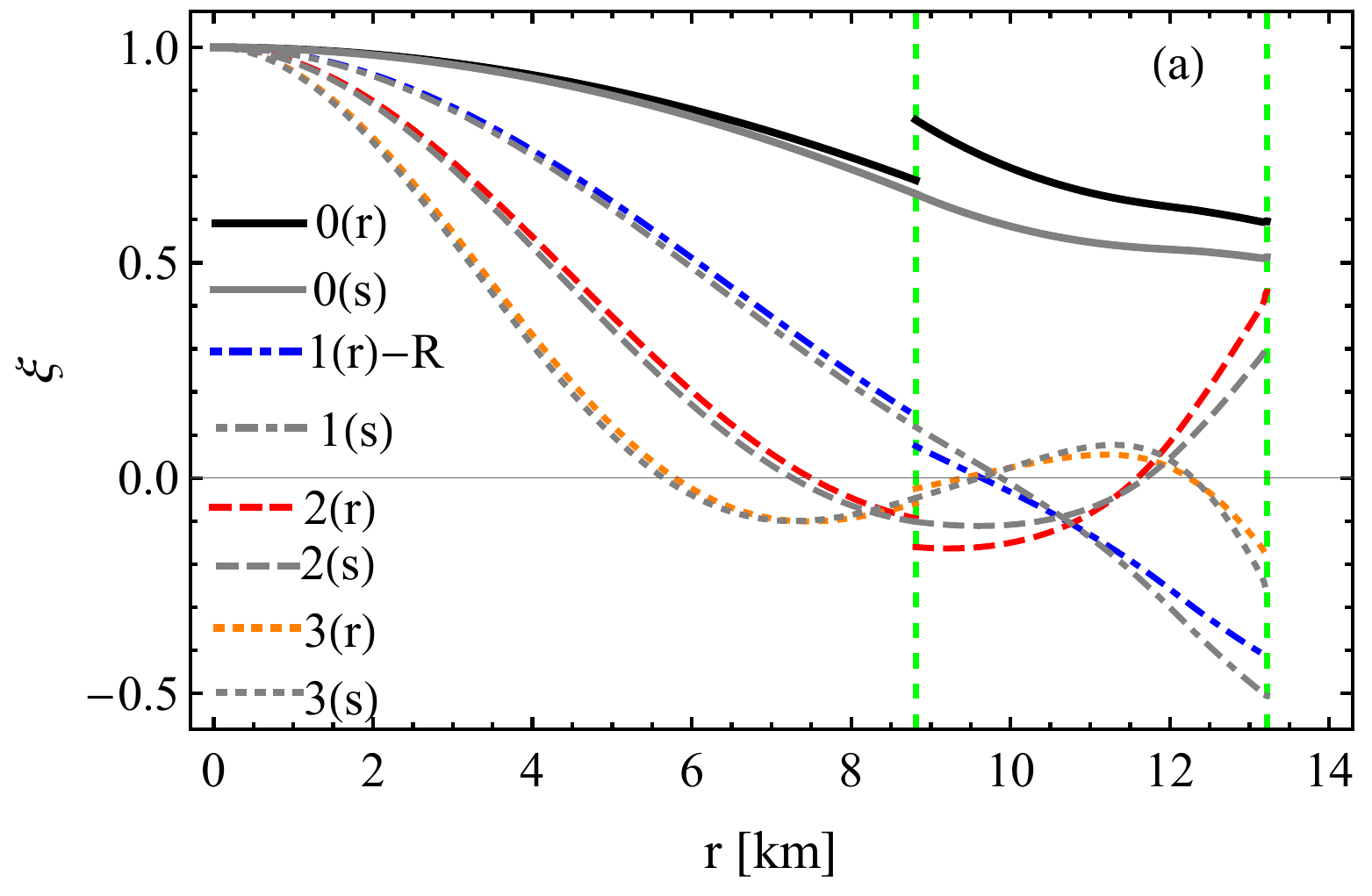}
    \includegraphics[width=\columnwidth]{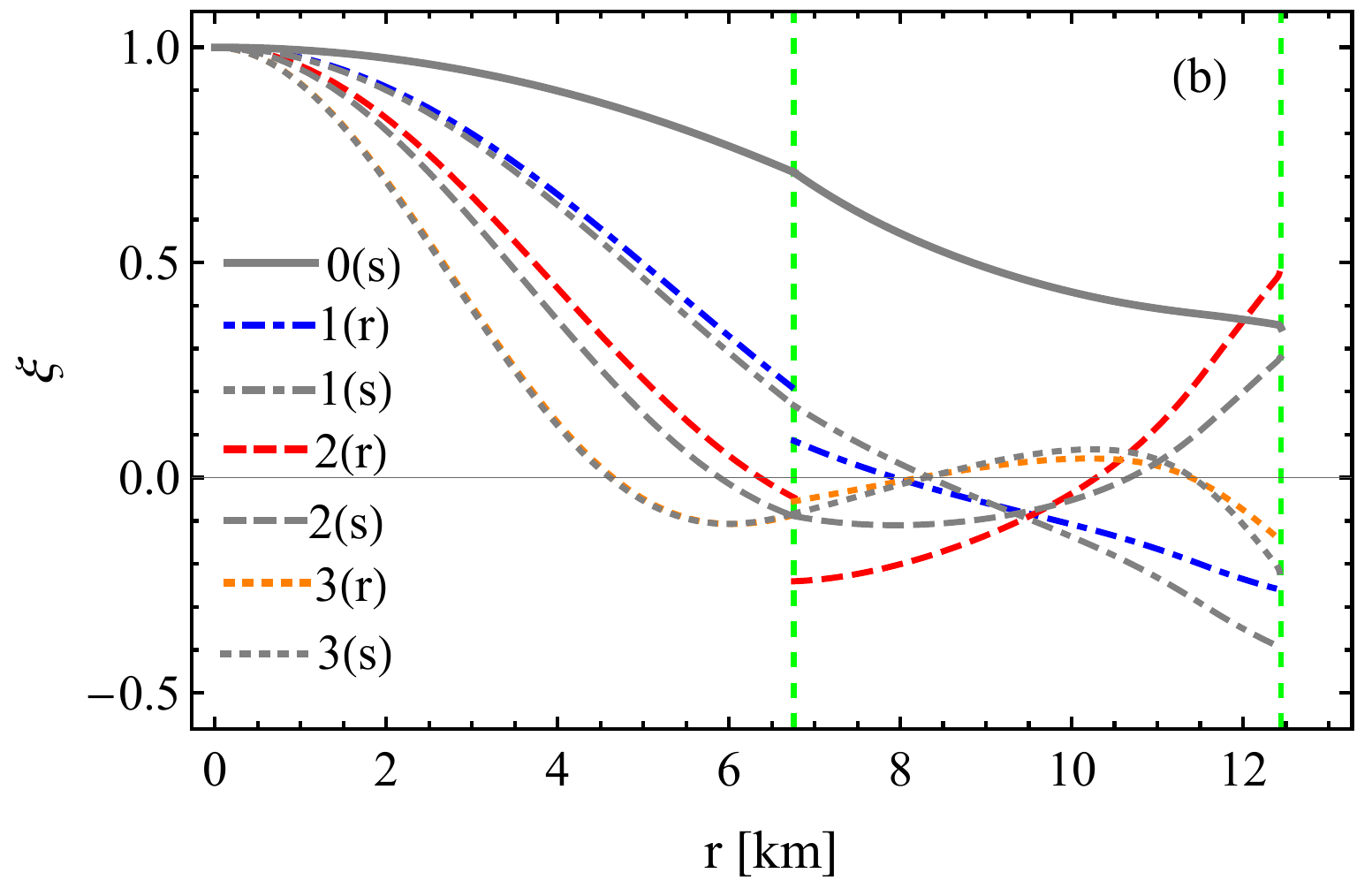}
    \caption{(a) First eigenmodes for the parameters in Fig. \ref{B_138_a4_0_5} with $\rho_c=11.05 \times 10^{14}$g.cm$^{-3}$, leading to $M=2.04$ $M_{\odot}$. In this case the phase transition radius is at $\sim 8.5$ Km. Note that the ordering of the modes is related to the number of nodes they exhibit.  (b) First excited eigenmodes for the parameters of Fig. \ref{B_145_a4_0_5} with $\rho_c=23.02\times 10^{14}$ g/cm$^{3}$, resulting in a star's mass of $M=2.13$ $M_{\odot}$ and phase transition radius of about $6.4$ Km. For this case the fundamental mode related to rapid phase transitions is nonexistent.  In both figures, one sees that the difference between slow and rapid eigenmodes tend to decrease for higher modes. }
\label{eigenmodes} 
\end{figure*}
%

\subsection{Eigenfunctions}

The left panel of Figure \ref{eigenmodes} shows the eigenfunctions $\xi(r)$ (for modes ranging from 0 to 3) using the same EOS parameters of Fig. \ref{B_138_a4_0_5} for a hybrid star with mass $2.04$ $M_{\odot}$. For this particular choice there are rapid and slow representatives for each mode depicted. Their order is defined by the number of nodes they present, even in the case where jumps take place. The fact that rapid phase transitions differ from slow ones only by a boundary condition does not mean that slow and rapid eigenmodes differ only after the phase transition radius. Rather, they should generally be different throughout the whole star because the description of perturbations constitutes a Sturm-Liouville problem, whose solutions are extremely dependent upon boundary conditions. However, broadly speaking, one sees from the figure that the higher the mode the smaller the jump for fast phase transitions, which make them approach the slow eigenmodes. 
The right panel of Figure \ref{eigenmodes} shows the eigenfunctions $\xi(r)$  using the same EOS parameters of Fig. \ref{B_145_a4_0_5} for the hybrid star with mass $2.13 M_{\odot}$. Note from it that for such a mass the rapid fundamental mode is nonexistent. The qualitative behavior of the eigenmodes in this case is similar to that of the left panel of Fig. \ref{eigenmodes}.

\subsection{Twin and triplet stars}

From the above analysis a very important conclusion ensues:  slow phase transitions lead to the existence of couples of stable stars with the same gravitational mass but different central densities,  with one of the stars having $\partial M / \partial \rho_c \leq 0$. We shall call these pair of stars as ``twins'', but keeping in mind that our twins are different in nature from the ones usually described in the literature, e.g., \cite{2017PhRvL.119p1104A,2015A&A...577A..40B,2017PhRvC..96d5809A}, given that the latter are configurations that always verify  $\partial M/\partial \rho_c>0$. Thus, in order to avoid ambiguity, we use in some cases the term ``slow twins''  for talking about a couple with one of the members having $\partial M / \partial \rho_c \leq 0$. Notice that for slow phase transitions one could even have three stable stars with the same mass (triplet) if the hadronic counterpart is also taken into account (see e.g. Fig. \ref{mass_densisty_B_160_a4_0_8}). In general, we could expect even multiplets if additional compact star branches were present.  Another conclusion that follows from the stability analysis of the previous subsections is that twin counterparts having $\partial M / \partial \rho_c \leq 0$ are not possible in the case of rapid phase transitions.

Another interesting feature that arises for some EOS parametrizations is the existence of some stable hybrid configurations with a gravitational mass smaller than the mass of the most massive one-phase (purely hadronic) object, i.e. the plateau mass in the $M-\rho_c$ plot. Such undercritical masses can be seen explicitly in Fig. \ref{mass_densisty_B_160_a4_0_8} for $\rho_c/\rho_{sat} \sim 2.5$ and in Fig. \ref{m180}  for $\rho_c/\rho_{sat} \gtrsim 5$ where it is apparent that for some central densities above the one where the hybrid phase starts to exist, stars could have undercritical masses if the transition is slow.

As a thumb rule, when critical points do not take place in the $(M,\rho_c)$ plot of hybrid systems, slow twin stars would be related to couples involving one hadronic star and a hybrid one (therefore utterly different systems with the same mass); when it only presents a local maximum, triplet stars (slow or otherwise) would not be possible and  twin stars could emerge just in the case of slow phase transitions.  If a third or even a fourth compact star family were present (as in e.g.  \citet{2017PhRvL.119p1104A}),  complex groups of multiplets could arise if the transition is slow. Nonetheless, in some cases only the existence of twin stars  might already talk about the kind of phase transition taking place inside neutron stars, as well as their hybrid nature.

\subsection{Density jumps for rapid phase transitions}

\begin{table*}
\centering
\begin{tabular}{@{}|c|c|cccc|@{}} 
\hline
$a_4$ & $B_{max}^{1/4}$~(MeV) & $M_{max}$~($M_{\odot}$) & $R_{M_{max}}$~(km) & $\eta_{B_{max}}~ (\epsilon_-/\epsilon_+)$ & $3/2(1+p_t/\epsilon_+)$ \\ \hline
0.40 & 168 & 2.75 & 13.51 & 2.01 & 2.15 \\
0.50 & 168 & 2.60 & 13.92 & 1.92 & 2.10 \\
0.55 & 167 & 2.47 & 14.06 & 1.86 & 1.94\\
0.60 & 166 & 2.29 & 14.14 & 1.81 &  1.87\\
0.65 & 164 & 2.01 & 14.11 & 1.75 & 1.79\\ \hline
0.70 & 164 & 1.71 & 11.65 & 1.76 & 1.73\\
0.75 & 168 & 1.63 & 11.33 & 1.89 & 1.72\\
0.80 & 171 & 1.55 & 10.70 & 2.00 & 1.71\\
0.85 & 176 & 1.50 & 11.87 & 2.17 & 1.72\\
0.90 & 178 & 1.44 & 10.42 & 2.27 & 1.70\\
0.99 & 185 & 1.35 & 10.10 & 2.54 & 1.71\\
\hline
\end{tabular} 
\caption{Maximum values of $B$ and masses ($M$) and associated value of $\eta$s and radii ($R$) for stable hybrid stars ($a_2^{1/2}=100$ MeV) when rapid phase transitions are taken into account.}\label{ta1}
\end{table*}

For specificity, we focus here only on hybrid stars with rapid phase transitions.
Naturally, for a given value of $a_4$, different values of $B$ lead to different maximum masses for the associated stars; see Fig. \ref{mmax}. 
However, there exists a maximum value of $B$ above which the fundamental mode becomes unstable (imaginary eigenfrequency). 
The change of the effective bag constant also changes the density jump at the phase transition radius, $\eta\equiv \epsilon_-/\epsilon_+$. For \textit{small} core radii, it is classically known that stable solutions should be related to $\eta < 3/2$ (see \cite{1981PhLB..101..366K} and references therein).  
When general relativistic corrections are taken into account, the above condition should be superseded by $\eta < 3/2(1+p_t/\epsilon_+)$ \citep{1971SvA....15..347S,1981PhLB..101..366K,1987A&A...172...95Z}. 
Given $B$, $a_4$, $a_2$ and a hadronic equation of state, one can find a unique transition pressure $p_t$ (related to the equality of the associated Gibbs functions of the quark and hadronic phases), which leads to 
\begin{equation}
\epsilon_-= 4B+3p_t+ \frac{3a_2}{4a_4\pi^2}\left[a_2+\sqrt{a_2^2+\frac{16}{3}a_4\pi^2(B+p_t)}\right]\label{rhominus}.
\end{equation}
From the hadronic equation of state $p_{ha}=p_{ha}(\epsilon_{ha})$, one can simply obtain $\epsilon_+$ by inverting it and evaluating $p_{ha}$ at $p_t$. 

Table \ref{ta1} summarizes maximum values of $B$ and associated stellar parameters for configurations with $0.3\leq a_4<1$ when rapid phase transitions are considered. ($B_{min}<B<B_{max}$ leads to $f_{0(r)}^2>0$ for some range of central densities.)  
The horizontal line in the middle of Table \ref{ta1} splits the (quark phase) parameters leading to stable stars with small quark cores (upper part) from those leading to 
stable stars with extended cores (lower part) [these parameters, however, lead to unstable stars when their cores are small, which happens when their central pressures are close to the associated $p_t$s].
As expected, one sees that the condition $\eta < 3/2(1+p_t/\epsilon_+)$ holds for stable stars with small cores.  Stable stars with extended cores, outside the scope of \cite{1971SvA....15..347S}, may have $\eta > 3/2(1+p_t/\epsilon_+)$. Actually, there are already known physically relevant cases where this should happen \citep{2008A&A...479..515Z,2013PhRvD..88h3013A,2016EPJA...52...62A,2015A&A...577A..40B,2017PhRvC..96d5809A} and in this sense the quark matter model we investigated is just another example thereof.

\section{Discussion and Conclusions}
\label{discussion}

It is known since long ago that in one-phase stars the stability analysis for radial oscillations is equivalent to the condition $\partial M/\partial \rho_c\geq 0$. This is very important because it gives a practical rule for determining whether or not a system lingers on in time when perturbed based solely on a sequence of its static solutions. However, when hybrid stars are taken into account, the aforementioned stability criterion should be taken cautiously. The reason is broadly due to the possibility of phase conversions and the non-differentiability of some physical quantities at the phase-splitting surface, such as the pressure and the energy density, related to the total different natures of the fluids in each phase. However, one is always on the safe side when the perturbation equations are directly analyzed, because the stability of a nonrotating system is related to the reality of the fundamental eigenfrequencies. The price to pay is naturally the dealing with nonlinear equations in the presence of nontrivial boundary conditions. Fortunately, powerful numerical tools already exist, allowing us to solve them relatively easily, such as the one we have made use of in this work.   

When extra boundary conditions are present, one expects the spectrum of eigenfrequencies and eigenmodes of the system to change, since the associated perturbation equations constitute a Sturm-Liouville problem. This, in turn, may in general lead the known conditions valid for one-phase systems to change. In particular, the practical rule that $\partial M/\partial \rho_c\geq 0$ for a system to be stable should not be taken for granted.

In the present paper, we have showed that within a phenomenologically inspired bag-like model for the quark phase and a nonlinear Walecka model for the hadronic phase (NL3 parametrization), the spectrum of hybrid stars in general relativity is extremely dependent upon the nature of the phase transitions taking place near the surface splitting the two aforesaid phases. 
When rapid phase transitions are present a new mode appears and it could in principle be any overtone. However, for the conditions studied in the present work (NS presenting first order phase transitions and with some solar masses) we have found the reaction mode to be either the fundamental mode or the first excited mode. We have also found that for rapid phase transitions the frequency of the fundamental mode is a real number if and only if the condition $\partial M/\partial \rho_c\geq 0$ is verified, coinciding with the practical rule for one-phase systems. 
Therefore, for rapid phase transitions, the above mentioned condition is necessary and sufficient for the stability of hybrid stars. 
When slow phase transitions are analyzed, though, $\partial M/\partial \rho_c\geq 0$ is neither necessary nor sufficient \textit{in general}, greatly contrasting with its one-phase counterpart. This shows that particular attention should be taken when analyzing the physical situations which lead to slow phase transitions.

We stress that our findings are basically related to the assumptions of cold and catalyzed matter even in the presence of perturbations and extra-boundary conditions to the perturbation eigenvalue problem. Cold and catalyzed matter can be justified when the timescales of weak reactions are much smaller than periods of oscillation. Extra-boundary conditions in our case encompass the physics of phase conversions, thus related to strong interactions. Due to the fact many that aspects influence the  matter conversion timescales, we just analyzed the possible extremes, i.e., when they are extremely low or extremely fast. Therefore, besides the timescale of perturbations, there are two other important (uncorrelated) timescales in problems of phase conversions, namely the weak interaction timescale and the time associated with the nucleation process of confinement and deconfinement. The possibility might also arise that for both rapid and slow phase transitions weak timescales are large when compared to the oscillations. In this case, one should take into account the adiabatic index of frozen compositions for the perturbation problem. This issue is nontrivial and we plan to carry it out elsewhere. It is already known, though, that when the frozen adiabatic index is considered instead of the equilibrium one the critical points of the $(M,\rho_c)$ plot do not coincide with null eigenfrequencies, but they appear at larger values of the central density, where $\partial M/\partial \rho_c<0$ \citep{1995A&A...294..747G}. One of the reasons for the violation of the theorems in \cite{Wheeler1965} is related to the fact that systems are not allowed to change composition upon density perturbations (not catalyzed) and a one-to-one correspondence between pressure and energy density is not guaranteed anymore (for further details, see \citet{1977ApJ...217..799C} and \citet{1972gcpa.book.....W}).

Qualitatively speaking, our results in the case of slow phase transitions lead to similar conclusions as the ones regarding frozen compositions, namely $\partial M/\partial \rho_c <0$ and still real fundamental eigenfrequencies. 
This is not surprising since slow boundary conditions embody the fact that volume elements upon perturbations are in a long lived metastable state (which is not the one with the lowest free energy) much in the same way as frozen fluid elements have a larger free energy than they would have if they were able to attain chemical equilibrium.
However, what is interesting about our analysis is that even when one assumes that just a portion of a system may convert from one phase to the other (just the vicinities of a phase-splitting surface), one could already have non-negligible consequences for it when compared to a one-phase star. The reason would be due to the fact that perturbations and eigenfrequencies have to greatly rearrange themselves in order to accommodate this transient region (similarly to what happens to a system of coupled springs when new masses are inserted).

In the context of the hadron and quark models we investigated, when rapid phase transitions take place, due to the condition $\partial M/\partial \rho_c\geq 0$, it would not be possible the occurrence of twin hybrid stars; see for instance the case related to Fig. \ref{mass_densisty_B_160_a4_0_8} and Sec. VII-C.
However, when slow phase transitions are taken into account, twin hybrid stars may emerge because some stars with $\partial M/\partial \rho_c\leq 0$ may have real eigenmodes in this case (see Figs. \ref{mass_densisty_B_160_a4_0_8} and \ref{B_160_a4_0_8}). 
If one also takes into account the sequence of hadronic stars, then slow phase transitions could even lead to the existence of triplet stars.  
For rapid phase transitions, though, only twin stars could emerge. All of the above shows that there may be in principle macroscopic ways of assessing internal processes of neutron stars and therefore the possibility of learning about its constitution. We leave for future work slow phase conversion investigations of hybrid models where {several families of compact} stars could already exist with $\partial M/\partial \rho_c>0$ (see for instance  \cite{2017PhRvL.119p1104A}).

The existence of a new branch of stable stellar configurations in the case of slow phase-conversions opens the possibility of several observable astrophysical phenomena. One is, as already mentioned, that it would be possible to detect twin or even triplet configurations having the same \textit{gravitational} mass $M$ but different radii.  On the other hand, interesting catastrophic scenarios may arise if there exist twin or triplet NS configurations having the same \textit{baryonic} mass $M_B$ but different radii. In such a case, the conversion of a hadronic star into a hybrid star with the same $M_B$ but a smaller $M$ might occur. In a similar way, a hybrid star with a small quark core could be converted into a hybrid star with a sizable quark core if the latter has the same $M_B$ but lower $M$. This kind of scenarios have been extensively analyzed in the literature (see e.g. \cite{Bombaci:2016xuj}, \cite{2008A&A...479..515Z} and references therein), and gains additional interest in the case of slow phase-conversions because in this case the existence of multiple stable configurations with the same $M_B$ is enhanced. 
For instance, consider the parameters of Fig. \ref{m180}, where hybrid stars could only exist if conversions are slow. (For mass-radius aspects of this case, see Fig. \ref{M-R-relation}.)
Define $M_{B}=4\pi m_b\int_0^R n_b(r)e^{\lambda(r)/2}r^2dr$, where $m_b$ is the mass of a baryon, and take it to be $m_{Fe}/56$ \citep{2008A&A...479..515Z}. Thus, for $M_B= 2.013 M_{\odot}$ ($\rho_c^{ha}=2.00 \rho_{sat}$ and $\rho_c^{hy}=13.86 \rho_{sat}$), it follows that $M_{ha}=1.667 M_{\odot}$ ($R=14.2$ km) and $M_{hy}=1.689M_{\odot}$ ($R=11.3$ km), which means that in this case energy should be given to the system to go from a purely hadronic to a hybrid configuration. However, for $M_B= 2.544 M_{\odot}$ ($\rho_c^{ha}=2.31 \rho_{sat}$ and $\rho_c^{hy}=6.91 \rho_{sat}$), one has that $M_{ha}=2.002 M_{\odot}$ ($R=14.4$ km) and $M_{hy}=1.997 M_{\odot}$ ($R=13.5$ km), and thus this hadronic star would release energy when turning hybrid.
Several mechanisms have already been suggested as possible triggers for the conversion, including stellar quakes, pulsar spindown, and delayed quantum nucleation of quark drops \citep{Bombaci:2006cs,Bombaci:2004mt,2008A&A...479..515Z}. Such conversions are expected to produce conspicuous emission of neutrinos, gravitational waves and gamma-ray bursts arising from the large scale rearrangement of the stellar configuration. A careful analysis of the consequences of these conversions is left for future work.

Our analyses have been performed in the context of non-rotating stars, where the issue of secular instabilities does not take place, but to date all known stars are known to rotate which in principle may induce such instabilities.
Notwithstanding, our analyses are relevant when viscous dissipation and nonlinear damping mechanisms suppress secular instability's exponential growth which is expected to be the case below some critical rotation rate in cold catalyzed NSs. 
In such circumstances dynamical stability determines NSs existence and stable configurations are characterized by $\omega^2>0$ (real eigenfrequencies) because it is only in this case that displacements of volume elements are bound (for radial displacements $\Delta r\propto e^{i\omega t}\xi(r) $). 
Hence this must be  equivalent to the minimum of the total energy in the equilibrium solution (from the TOV equations) for fixed baryon number since it leads to the same conclusion regarding the dynamics of small volume displacements. 
We stress that in the presence of perturbations the total energy of the system changes with respect to the background to second order in $\xi$ \citep{Wheeler1965}. This is the case exactly because the TOV equations extremize the total mass \citep{Wheeler1965}. Therefore, each point in the $(M-\rho_c)$ plot is a \textit{local extreme} to the total energy. Stability arises if it is a \textit{local minimum} with respect to the total energy. This is exactly determined by the reality of the fundamental eigenfrequency, as explained previously.

It has been speculated that hybrid stars may contain a mixed hadron-quark phase in their interiors. In such a phase it is assumed that the electric charge is zero globally but not locally, and therefore charged hadronic and quark matter may share a common lepton background, leading to a quark-hadron mixture extending over a wide density region of the star \citep{Glendenning:2001pe}. The mixed phase entails a smooth variation of the energy density, leading in turn to a continuous density profile along the star. 
Whether the quark-hadron interface is actually a sharp discontinuity or a wide mixed region depends crucially on the amount of electrostatic and surface energy needed for the formation of the varying geometric structures of one phase embedded in the other all along the mixed phase \citep{Voskresensky:2002hu,Endo:2011em,Yasutake:2014oxa}.
If the energy cost of Coulomb and surface effects exceeds the gain in bulk energy, the scenario involving a sharp interface turns out to be favorable. However,  there are still many points to be elucidated before arriving to an undisputed description of the quark-hadron coexistence. In particular, model calculations of the surface tension span a wide range of values (see \cite{Lugones:2013ema,Lugones:2016ytl} and references therein) and as a consequence the very nature of the quark-hadron interface remains uncertain.
Nonetheless, since in the presence of the mixed phase all physical quantities are differentiable everywhere,  these stars behave essentially as one-phase objects whose stability is not affected by the junction conditions explored in the present work. This fully justifies our assumption of disregarding it in our analysis.

Besides twin stars, one could in principle unveil the nature of phase transitions and innermost phases in neutron stars by means of direct frequency observations. This would especially be so due to the fact the reaction mode is uniquely associated with rapid phase transitions, it is very sensitive to the quark EOS and in some cases, such as when it is the fundamental mode, it may differ significantly from its slow counterpart for some masses (see, e.g., Fig. \ref{B_145_a4_0_5}). 
Such observations might be possible in the future through gravitational wave detectors because in the case of rotating objects we can expect some amount of gravitational radiation from even the lowest ($l = 0$) quasiradial mode \citep{Stergioulas:2003yp,Passamonti:2005cz}.  Additionally, purely radial oscillations could leave some electromagnetic imprint in the microstucture of pulsar emission or in magnetar flare lightcurves.

Summing up, in this work we have found the extra boundary conditions appropriate for slow and rapid phase transitions in hybrid stars. By using a bag-like model for the quark phase and a relativistic mean field theory model for the hadronic phase of a star, we have obtained that rapid phase transitions lead the reaction mode to be just a generalization of its classical counterpart. 
The general relativistic stability jump condition for hybrid stars with small cores $\epsilon_-/\epsilon_+ < 3/2(1+p_t/\epsilon_+)$ (Seidov's condition) holds true for rapid phase transitions, while it may not be valid for large (extended) cores.
For hybrid stars experiencing rapid phase transitions the frequency of the fundamental mode is zero when $\partial M/\partial \rho_c=0$, identical to what is known for one-phase stars. However, slow phase transitions lead to $\partial M/\partial \rho_c < 0$ for the stellar configuration at which the frequency of the fundamental mode is zero. 
As a consequence, the range of central densities in  hybrid stars where slow phase transitions take place at the interface is larger than in hybrid objects undergoing rapid phase transitions.
For the hadronic and quark models studied here, twin hybrid stars could exist for slow phase transitions, while this would not be the case for rapid ones. When one-phase hadronic stars are also taken into account, even triplet stars could emerge for slow phase transitions.  
For other sets of EOSs leading to a third or a fourth compact star family,  complex groups of multiplets might arise if the transition is slow. In principle, it looks that many internal aspects of compact stars could be probed with these systems.

\acknowledgments
We are thankful to the anonymous referee for the raise of very good questions and Prof. Nils Andersson for insightful discussions. J.P.P. acknowledges the financial support given by Funda\c c\~ao de Amparo \`a Pesquisa do Estado de S\~ao Paulo (FAPESP) under grants No. 2015/04174-9 and ~2017/21384-~2.  C.V.F is likewise grateful to Comiss\~ao de Aperfei\c coamento de Pessoal do N\'ivel Superior (CAPES) of the Brazilian government. G.L. acknowledges the Brazilian agencies Conselho Nacional de Desenvolvimento Cient\'{\i}fico e Tecnol\'ogico (CNPq) and FAPESP for financial support.

\bibliography{fast_radial_transitions}
\bibliographystyle{apj}

\end{document}